\documentclass[
    10pt
    ,aps
    ,pop
    ,oneside
    ,nofootinbib
    ,floatfix
    ,twocolumn
    ,nolongbibliography
    ,nolinenumbers
    ,showkeys
    ,noeprint
]{revtex4-1}

\usepackage{graphicx}
\usepackage[
    colorlinks
        ,linkcolor=violet
        ,citecolor=teal
    ,unicode
    ,pdfpagemode=UseOutlines
    ,pdfdisplaydoctitle=true
    ,pdfpagetransition=Wipe
    ,pdfusetitle
]{hyperref}
    \hypersetup{
        ,pdfkeywords={GDMT, diamagnetic bubble, finite Larmor effects, adiabatic invariant}
    }

\usepackage{bookmark}

\usepackage{amsmath,amssymb,bm}
\allowdisplaybreaks[1]
\usepackage{siunitx}
\usepackage{ifluatex,ifxetex}
\ifnum 0\ifxetex 1\fi\ifluatex 1\fi>0
    \usepackage{fontspec}
    \defaultfontfeatures{Ligatures={TeX},Renderer=Basic}

    \usepackage[mathrm=sym,math-style=ISO,bold-style=ISO]{unicode-math}

\else

    \usepackage[T1]{fontenc} 
    \usepackage[utf8]{inputenc} 
    \usepackage{newtxtext,newtxmath}
\fi

\usepackage[russian, english, showlanguages]{babel}

    \ifnum 0\ifxetex 1\fi\ifluatex 1\fi>0
        \renewcommand{\vec}[1]{\symbf{#1}}
    \else
        \renewcommand{\vec}[1]{\mathbf{#1}}
    \fi

    \DeclareMathOperator{\Div}{div}

    \DeclareMathOperator{\const}{const}

    \newcommand{\Heviside}{H}
    \newcommand{\Hamiltonian}{\mathcal{H}}

    \DeclareMathOperator{\curl}{curl}
    \DeclareMathOperator{\sign}{sign}

    \newcommand{\tparder}[2]{{\partial #1}/{\partial #2}}
    \newcommand{\parder}[2]{\frac{\partial #1}{\partial #2}}

    \newcommand{\dif}[2][]{\mathop{}\!\mathrm{d}
        \if
            \relax\detokenize{#1}\relax
        \else
            ^{\mkern-1.mu#1}\mkern-2.5mu 
    \fi
    #2}

    \usepackage[svgnames]{xcolor}
    \definecolor{darkblue}{cmyk}{1.00, 0.50, 0.00, 0.40}

\begin{document}

\def\bot{\mathrel\perp}

\title{On the structure of the boundary layer in a Beklemishev diamagnetic bubble}
\author{Igor Kotelnikov}
    \email{I.A.Kotelnikov@inp.nsk.su}
    \affiliation{Budker Institute of Nuclear Physics, Novosibirsk, Russia}
    \affiliation{Novosibirsk State University, Novosibirsk, Russia}


\begin{abstract}

    The article provides a kinetic description of the plasma equilibrium in the Beklemishev diamagnetic trap, where the traditional approach based on the theory of magnetic drifts is not applicable, since the ions move in a substantially non-circular orbit, the diameter of which is approximately equal to the diameter of the diamagnetic bubble. The ion distribution function was found in the collisionless approximation, neglecting the diamagnetic electron current. The radial profile of the magnetic field, the plasma density, the current density, and the components of the pressure tensor are calculated. It was found that the width of the boundary layer in the diamagnetic bubble varies from 6 to 8 Larmor radii calculated by the vacuum magnetic field. An adiabatic invariant is calculated that replaces the magnetic moment, which is not conserved in the diamagnetic bubble. The criterion of absolute confinement is formulated and the plasma equilibrium is found for the case when the adiabatic invariant is not conserved and only ions whose velocity satisfies the criterion of absolute confinement are trapped.
\end{abstract}

\keywords{Beklemishev diamagnetic bubble; gas-dynamic trap; adiabatic invariant; absolute confinement}

\maketitle

\section{Introduction}\label{s1}


The renewed interest in linear traps for plasma confinement, also called open or mirror traps, has recently been associated with the proposal of Alexei Beklemishev to form a diamagnetic ``bubble'' \cite{Beklemishev2016PoP_23_082506} in such a trap. Beklemishev's bubble (if its feasibility will be proved) dramatically increases the chances of open traps for the role of a thermonuclear reactor. According to Beklemishev himself, the diamagnetic bubble is halfway between the field reversed configurations (FRC) and the linear gas-dynamic trap (GDT).

\begin{figure}[pb]
  \centering
  \includegraphics[width=\linewidth]{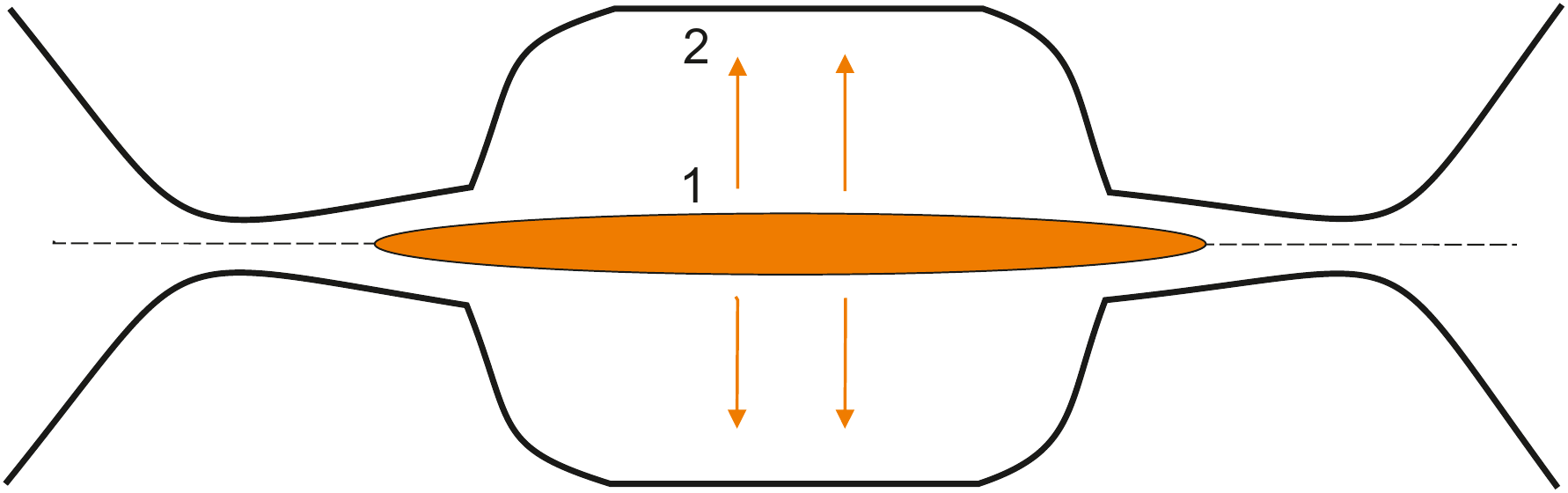}
  \caption{
    Expansion of initially thin flux tube \emph{1} at high $\beta$ leads to corresponding increase in the effective mirror ratio of a linear trap. If there is a quasi-uniform patch of the vacuum magnetic field at the bottom of the magnetic well, the resulting “diamagnetic bubble” \emph{2} will be roughly cylindrical. The plasma boundary at cylinder ends needs MHD stabilization (not shown).
  }\label{fig:Beklemishev2016PoP_23_082506_01}
\end{figure}


As shown in \cite{Beklemishev2016PoP_23_082506}, the plasma equilibrium in a linear trap at $\beta \approx 1 $ (above the mirror instability threshold), turns into a kind of diamagnetic “bubble” while maintaining the topology of the magnetic field. It can take two forms: either the plasma volume expands significantly in the radial direction, while retaining the trapped magnetic flux, as is schematically shown in Fig.~\ref{fig:Beklemishev2016PoP_23_082506_01}, or the distribution of the plasma over the magnetic flux tubes changes if the plasma radius is limited by a diaphragm so that the same cross section contains a significantly reduced magnetic flux. If the magnetic field of the trap is quasihomogeneous around its minimum, the bubble can be made approximately cylindrical with a radius much larger than the radius of the corresponding vacuum flux tube and with nonparaxial ends. Then the effective mirror ratio of the diamagnetic trap becomes very large, but the transverse plasma cross-field transport increases.

The idea of the Beklemishev diamagnetic bubble has some backstory. Probably Ilya Lansky was the first \cite{Lansky1993BINP_96} to clearly state that a violation of the conditions
    \begin{equation}
    \label{1:11}
    \parder{}{B}
    \left(
        \frac{B^{2}}{8\pi} + p_{\bot}
    \right)
    > 0
    ,
    \end{equation}
%
which in a homogeneous plasma guarantees the absence of mirror instability, in an axisymmetric open trap is accompanied by a violation of the paraxial approximation. However, it is possible that this fact was known by William Newcomb, who in his review article ``Equilibrium and stability of collisionless systems in the paraxial limit'' \cite{Newcomb1981JPP_26_529}  indicated the inequality \eqref{1:11} as the one of the two conditions for the applicability of his theory. Newcomb refers to a large article by Harold Grad \cite{Grad1967PSAM_18_167}, which lists the condition \eqref{1:11} among the sufficient conditions for plasma stability, which he calls well-posedness creteria. Lansky quotes Grad's work, but does not mention Newcomb's article. For a model distribution function, he calculated the limiting value of the relative plasma pressure $\beta$, above which the paraxial approximation is violated and large gradients of the magnetic field, plasma density and pressure appear.


The next step was made by Konstantin Lotov \cite{Lotov1996PoP_3_1472}. He has  shown that in open field line geometry with high expansion ratios of the flux conserver (superconducting expander) there appears a region of zero magnetic field near the device axis.
Lotov's solution was the first, although somewhat artificial, example of diamagnetic cavity as a zero-field region inside the mirror-trap plasma.
%
Note that Lotov does not mention the work of Grad, Newcomb and Lansky.

More realistic example of diamagnetic ``hole'' was given by Igor Kotelnikov \cite{Kotelnikov+2010PhysRevE_81_067402, Kotelnikov2011FST_59_47}.
Within the framework of paraxial approximation he has shown that in an anisotropic plasma with sloshing ions confined an open-ended system a magnetic hole is formed near the turning point of the sloshing ions above the threshold of the mirror instability. The magnetic field experiences a jump at the hole boundary from the side of the magnetic mirror. For a small excess over the mirror instability threshold, the surface of the discontinuity has the shape of a truncated paraboloid, and the magnitude of the magnetic field jump at the system axis is proportional to the radius of the hole and gradually decreases to zero away of the axis. It was argued that disappearance of the magnetic hole because of the widening of the sloshing ions angular spread in the course of the neutral beam injection results in abrupt anticorrelated changes of the diamagnetic signals measured near the turning point of the sloshing ions and near the midplane of the gas-dynamic trap as was confirmed experimentally on the gas-dynamic trap.
%
%
%
Kotelnikov in his papers \cite{Kotelnikov+2010PhysRevE_81_067402, Kotelnikov2011FST_59_47} mentions Lansky's preprint \cite{Lansky1993BINP_96}, papers by Lotov \cite {Lotov1996PoP_3_1472} and Grad \cite{Grad1967PSAM_18_167}. In turn, Beklemishev in \cite{Beklemishev2016PoP_23_082506} several times cites the papers by Kotelnikov but ignores other predecessors.


Stationary diamagnetic cavities were discovered in the solar wind in the shadow of various objects, such as the Moon \cite{Colburn+1967Science_158_1040} and Comet of Churyumov-Gerasimenko \cite{Goetz+2016MNRAS_462_459, Goetz+2016AA+588_26, Nemeth+2016MNRAS_462_415}. Numerous authors performed experimental \cite{Zakharov+1999_JPlasFusRes_2_398, Bamford+2008PPCF_50_124025} and theoretical works \cite{Nykyri+2011JGRSPh_116_A06208, Thomas+1988JGRSPh_93_11341, Thomas1988JGRSPh_94_13579, Kuznetsov+2015PoP_22_042114} on simulation of stationary and emerging diamagnetic cavities in the solar wind. Although the study of diamagnetic cavities in space plasma constitutes a significant and independent line of research, it seems that the results obtained there do not find so far direct application in plasma physics of open traps because of disparate plasma parameters.


The confinement time in the Beklemishev bubble
    \begin{equation}
    \label{1:01}
    \tau_{n} \approx \sqrt{2\tau_{\|}\tau_{\bot}}
    \end{equation}
was evaluated in \cite{Beklemishev2016PoP_23_082506} from a solution of the system of magnetohydrodynamic and transport equation.
%
%
In the gas-dynamic (GD) approximation, the longitudinal confinement time of a plasma with sound velocity $c_{s}$ in a trap with the mirror ratio $ R $ and length $ L $ is estimated as
    \begin{equation}
    \label{1:02}
    \tau_{\|} = \tau_{\textnormal{GD}} = \frac{LR}{2c_{s}}
    .
    \end{equation}
The time of transverse diffusion in a plasma with a radius $ a $ in \cite{Beklemishev2016PoP_23_082506} was calculated in the approximation of resistive magnetohydrodynamic by the formula
    \begin{equation}
    \label{1:03}
    \tau_{\bot}
    =
    \frac{4\pi\sigma a^{2}}{c^{2}}
    ,
    \end{equation}
where $\sigma$ denotes the effective transverse conductivity of plasma.
This estimate in principle allows construction of relatively compact fusion reactors with lengths in the range of a few tens of meters.

Inside the diamagnetic bubble, the magnetic field is close to zero, but rapidly grows in the boundary layer. The characteristic radial scale $\lambda $ in the approximation of resistive magnetic hydrodynamics is written in \cite{Beklemishev2016PoP_23_082506} as
    \begin{equation}
    \label{1:04}
    \lambda = \sqrt{
        \frac{c^{2}}{4\pi\sigma }
        \frac{LR}{4c_{s}}
    }
    .
    \end{equation}
Consequently, the plasma lifetime in the diamagnetic bubble can also be written as
    \begin{equation}
    \label{1:05}
    \tau_{n} \approx \frac{a}{\lambda } \,\tau_{\|}
    .
    \end{equation}
%
It follows from this estimation that if the thickness of the boundary layer $ \lambda $ exceeds the value \eqref{1:04} predicted by the resistive MHD, then the initial estimates of the fusion prospects of the diamagnetic bubble may turn out to be overly optimistic. In particular, with the plasma parameters cited in \cite{Beklemishev2016PoP_23_082506}, the Larmor radius of ions is approximately an order of magnitude greater than $ \lambda $.
%
%


Calculations of the magnetic field in a diamagnetic bubble based on MHD approach are published in \cite{KhristoBeklemishev2019PlasFusRes_14_2403007}, where the numerical model of the axially symmetric equilibrium was based on the
coupled Grad-Shafranov and transport equations. However, as mentioned in \cite{Beklemishev2016PoP_23_082506}, the MHD approximation is not quite applicable. Indeed, the trajectories of ions in reality may be extended far beyond the predicted thin MHD-boundary. Even ions passing right through the middle of the “bubble” will reflect back only after passing one Larmor radius into the magnetic field of the border, making its width of the order of $\rho_{i} \sim c/\omega_{pi}$.
%
%
It was mention therefor in \cite{Beklemishev2016PoP_23_082506} that \emph{``the qualitative and quantitative descriptions of kinetic processes within the `bubble' boundary are definitely very important and should be addressed in the near future.''}

The outer magnetic configuration of a linear trap in the “bubble” regime is equivalent to the FRC scrape-off layer, though the FRC itself is replaced by the low-field “bubble.”
%
%
There are a large number of publications devoted to calculations of plasma equilibrium in FRC. In particular, in a series of papers \cite{RostokerQerushi2002PoP_9_3057, QerushiRostoker2002PoP_9_3068, QerushiRostoker2002PoP_9_5001, QerushiRostoker2003PoP_10_737} Rostoker and Qerushi examined plasma equilibrium in FRC with one- and two-dimensional approximation assuming rigid-rotation ion and electron distributions.
%


Our first goal in the kinetic description of the diamagnetic bubble will be the classification of ion trajectories with a large Larmor radius in Section \ref{s2}. One motivation for this study comes from the fact that some ions pass through regions of vanishing magnetic field, invalidating the conventional drift approximation. For the FRC case, such a classification was made in \cite{Wang+1979NF_19_39}. Section \ref{s2} concludes with a discussion of the ion distribution function $ \overline {f} $ in the configuration space of noncanonical variables $(v, \alpha; r, \psi) $.

In Section \ref{s3}, macroscopic quantities are calculated, such as plasma density, current density, and pressure tensor in the boundary layer in the one-dimensional approximation based on the analysis of ion trajectories assuming the boundary layer can be considered flat.

The magnetic field profile in the plane boundary layer is found in Section \ref{s4}. In Section \ref{s5} the distribution function $ f $ of ions in the phase space of the canonical variables $ (v, \theta; r, \psi) $ is written for the plane-boundary-layer approximation and it is shown that the calculation of its moments allows reproducing and supplementing the results of Section \ref{s3} in a simpler way.

In Section \ref{s6}, the theory developed for a flat boundary layer is generalized to the two-dimensional case; here, the profile of the magnetic field in the cylindrical diamagnetic bubble is calculated.

In Section \ref{s7}, an adiabatic invariant is calculated that replaces the magnetic moment, which is not conserved in the diamagnetic bubble, since the diameter of the trajectory of the charged particle is not small compared with the diameter of the diamagnetic bubble. Section \ref{s8} discusses absolute confinement criteria; it is shown here that a magnetic field penetrates into a diamagnetic bubble if the magnetic moment is not conserved and only ions in the absolute confinement region are trapped.

Finally, in Section \ref{s99} the main results and conclusions are formulated. Throughout this paper, we neglect the contribution of electrons to the diamagnetic current, assuming without further explanation that their temperature is quite low.

\section{
    Ion trajectories in a diamagnetic bubble
}
\label{s2}


Let us consider a model of a diamagnetic bubble in the form of an infinite cylinder of radius $ a $, inside which the magnetic field is zero, but for $ r> a $ in the boundary layer of thickness $ \lambda $ it increases to a sufficiently large vacuum field $ B_{\textnormal{v}}$. Then we can assume that the boundary layer reflects all the ions in the cylinder back to the interior of the cylinder in the same way that a mirror reflects a ray of light. The movement of ions along the axis of the cylinder does not interest us so far, therefore we will assume that the ion trajectory lies in the plane $ z = \const $ perpendicular to the axis of the cylinder.

\begin{figure}
  \centering
  \includegraphics[width=0.47\linewidth]{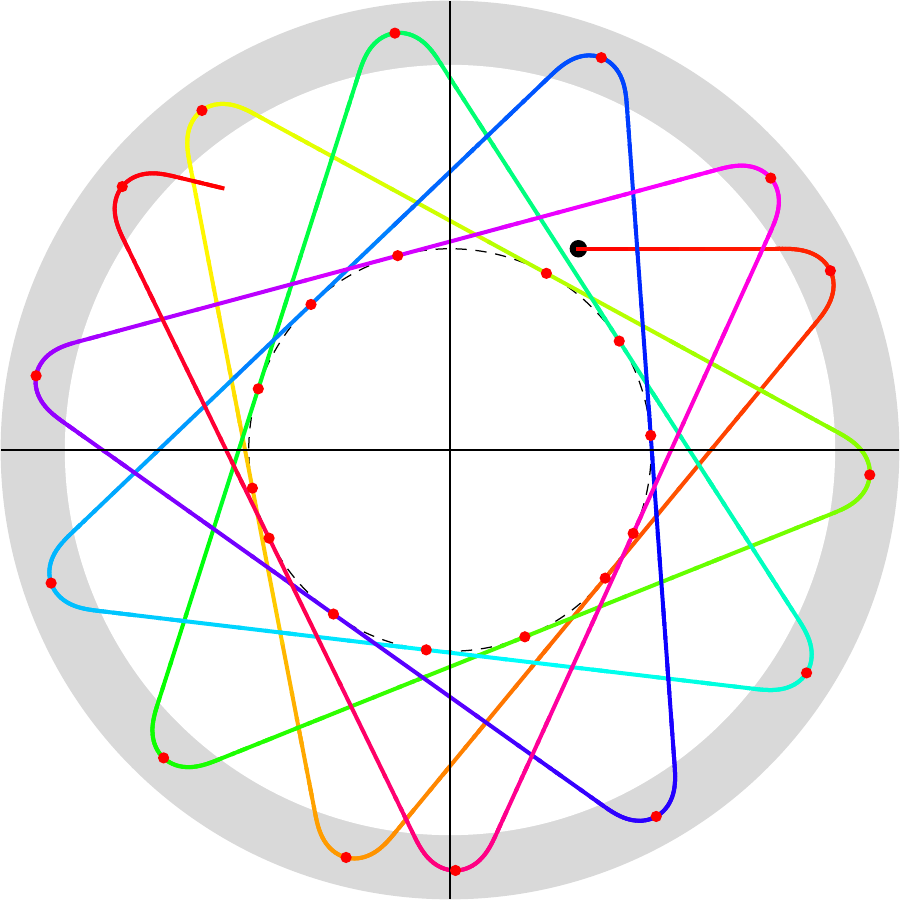}\hfil
  \includegraphics[width=0.47\linewidth]{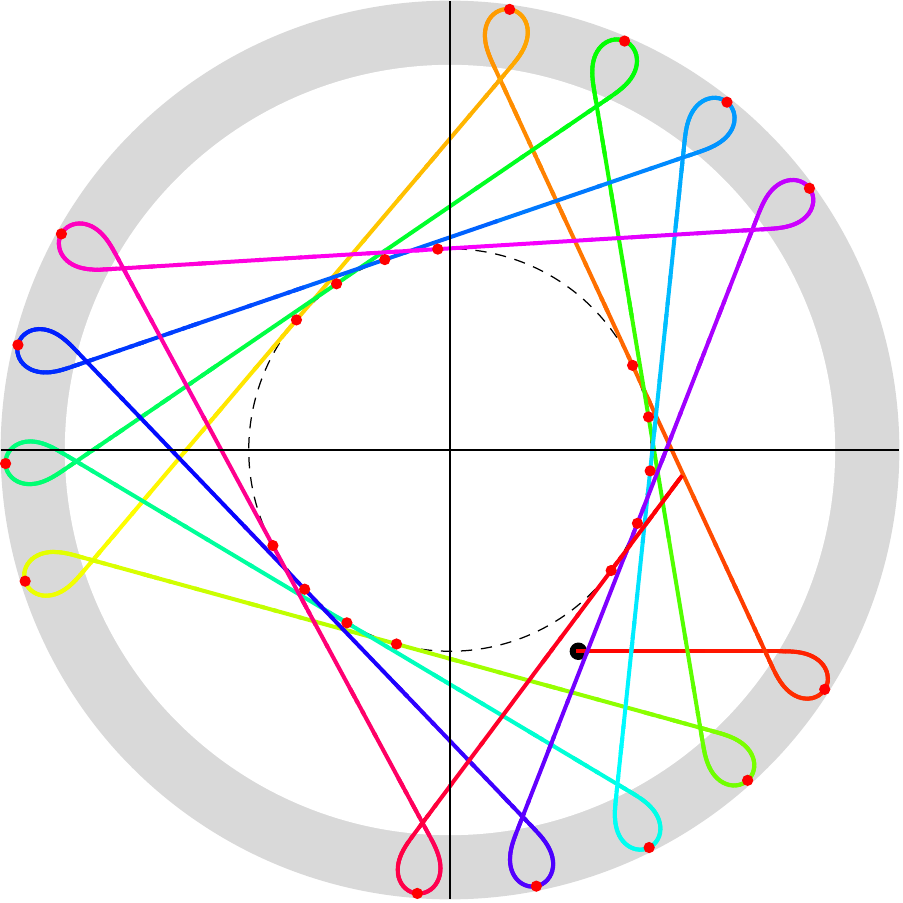}\\
  \includegraphics[width=0.47\linewidth]{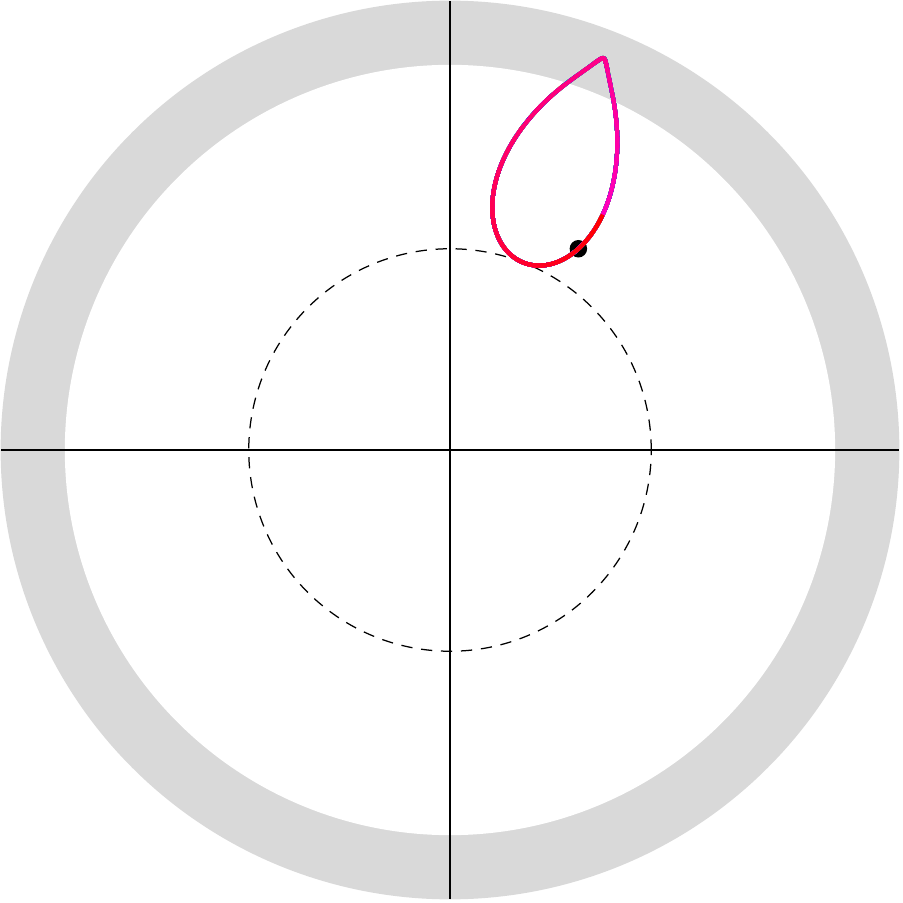}\hfil
  \includegraphics[width=0.47\linewidth]{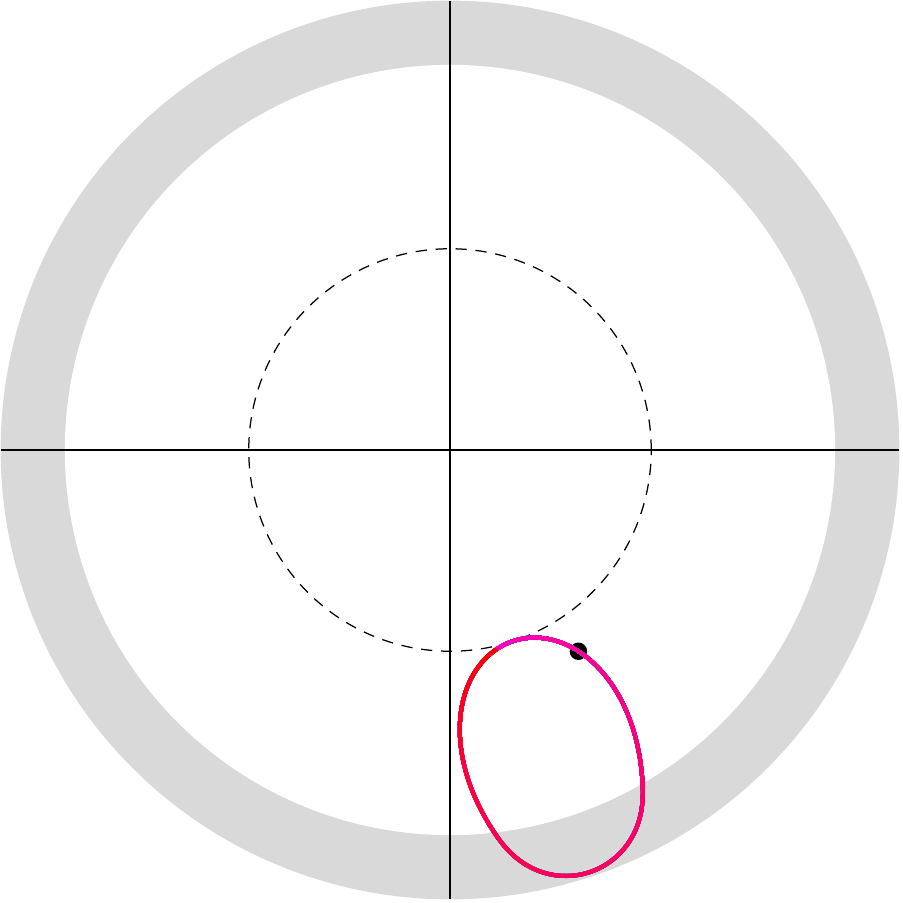}\\
  \caption{
    %
    The trajectory of the ion in the diamagnetic bubble at $\alpha = -0.35\pi/2$ (top left) and $\alpha = + 0.35\pi/2 $ (top right). The start point of the trajectory is marked with a black dot. The boundary layer, in which the magnetic field gradually increases from zero to the magnitude of the vacuum field, is shown in gray. In the figure on the left is $\Omega p_{\psi}<0 $ and the ion path bypasses the axis of the bubble in the direction of cyclotron rotation (clockwise), in the figure on the right $\Omega p_{\psi}> 0 $ and the path bypasses the axis against cyclotron rotation (counterclockwise). The bottom row shows the same trajectories in a rotating coordinate system with a specially tuned rotation frequency.
  }\label{fig:MotionInCylinder}
\end{figure}
%
Inside the cylinder, in the complete absence of a magnetic field, ions move along straight lines, as shown in Fig.~\ref{fig:MotionInCylinder}. In the boundary layer, which is shaded in gray in the figure, the magnetic field gradually increases from zero to $ B_{\textnormal{v}} $. We denote the angle of incidence of the ion on the boundary layer by the Greek letter $\alpha $ and agree that the angle $\alpha $ is counted from the normal to the boundary, as in the problem of the incidence of a light wave at the interface between two media. The direction of the normal coincides with the direction of the radius drawn from the axis of the cylinder to the point of incidence of the trajectory on the inner side of the boundary layer. As in optics, the angle of incidence is equal to the angle of reflection. The angle $\alpha $ can vary from $ -\pi/2 $ to $ +\pi/2 $. Let us agree that the negative values of $\alpha $ correspond to the case when the tangent to the boundary component of the ion velocity is directed towards the cyclotron rotation of the ion, as shown on the left in Fig.~\ref{fig:MotionInCylinder}. Then the ion trajectory bypasses the axis of the cylinder in the direction of cyclotron rotation, i.e.,\ clockwise. In the case of $\alpha > 0 $, the ion path bypasses the cylinder axis in the direction opposite to the direction of cyclotron rotation, as shown on the right in Fig.~\ref{fig:MotionInCylinder}. For some discrete values of the angle $\alpha $, the ion trajectory may become closed after several rounds around the cylinder axis, but in the general case it densely covers a ring with an inner radius
    \begin{equation}
    \label{2:01}
    b=a\left|\sin\alpha\right|,
    \end{equation}
where $ a $ denotes the inner radius of the boundary layer.


Since the angle of incidence is a conserved quantity, it (like the radius $ b $) is a convenient parameter in describing the distribution of ions. In the appendix \ref{A1}, it is proved that the angle $\alpha $ is preserved not only in a cylinder with a constant radius $ a $, but also with an adiabatically slow change in $ a $. We will find the ion distribution function inside the cylinder, assuming first that all ions are characterized by a single value of the angle $\alpha $ and the same velocity $ v $. Following Landau and Lifshitz \cite{LLX(eng)} (see also \cite[\S10.1]{Kotelnikov2013Binom(eng)}), we distinguish the distribution function $ f $ defined as the density of particles in the \emph{phase} space of canonically conjugate coordinates and momenta from distribution functions $\overline {f} $ defined as the particle density in the \emph{configuration} space of noncanonical variables. Generally speaking, the function $ f $ can be expressed in terms of noncanonical momenta and coordinates, preserving its meaning of particle density in phase space. Moreover, it does not become identically equal to the function $ \overline {f} $, since $ \overline {f} = Jf$, where $ J $ is the Jacobian of the transformation to a noncanonical set of coordinate and momentum.

\begin{figure}
  \centering
  \includegraphics[width=\linewidth]{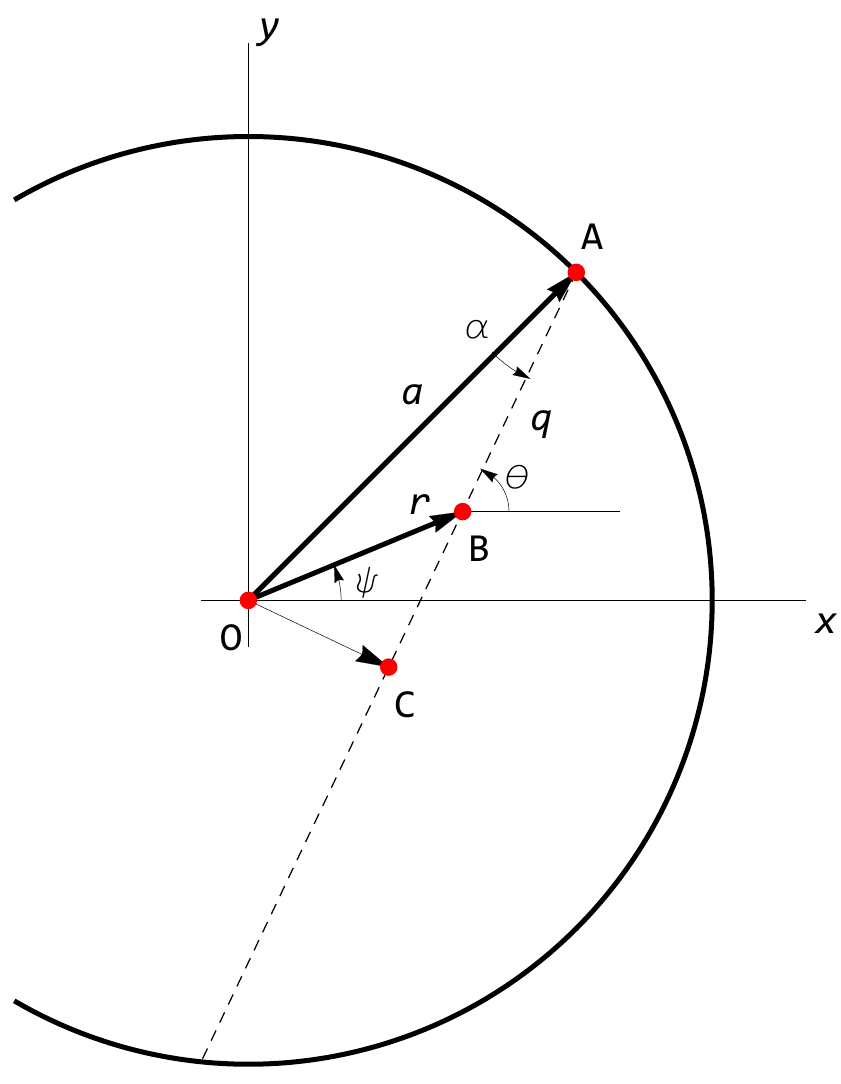}
  \caption{
    %
    The coordinate system in the diamagnetic bubble: the trajectory of the ion between the reflection points at the boundary is shown by a dashed line; the angle $\theta $ defines the slope of the velocity vector to the axis $ x $ at the point with polar coordinates $ (r,\psi) $; the angle of incidence $\alpha $ of the ion on the bubble boundary is measured from the normal to the boundary. The angle $\alpha $ can be expressed in terms of $\theta $ and $\psi $ by composing Eqs.~\eqref{2:12} for the triangle OAB, which, after excluding parameter $ q $, lead to Eq.~\eqref{2:02a}. In another way, Eq.~\eqref{2:02a} is obtained from the equality of the common leg OC in the triangles OAC and OBC.
  }\label{fig:CoordinatesInCylinder}
\end{figure}

%
The ion distribution function on a separate straight segment of the trajectory between successive reflection points at the cylinder boundary can be written as
    \begin{equation}
    \label{2:02}
    \overline{f}(v,\alpha; r,\psi) = A(v,\alpha)\,\delta (a\sin(\alpha)-r\sin(\theta -\psi))
    ,
    \end{equation}
%
assuming that the indicated segment is shown by a dashed line in Fig.~\ref{fig:CoordinatesInCylinder}, makes the angle $\theta$ with the axis $x$ and passes through the point With cylindrical coordinates $(r,\psi)$ at a distance $ b = a\left|\sin\alpha\right|$ from the axis O of the cylinder. This representation expresses the fact that the particle density at each point of the segment defined by the equation
    \begin{equation}
    \label{2:02a}
    a \sin(\alpha ) = r \sin(\theta - \psi)
    \end{equation}
%
is the same. The normalization factor $ A(v,\alpha) $ will be determined later. Further, it should be noted that there is a continuous set of such segments, as shown in Fig.~\ref{fig:MotionInCylinder}. It corresponds to the rotation of the picture around the axis O by an arbitrary angle $\psi_{0} $. Integrating over $\psi_{0} $, we get
    \begin{equation}
    \label{2:03}
    \overline{f}(v,\alpha; r)
    =
    \int_{0}^{2\pi}
    \overline{f}(v,\alpha; r,\psi-\psi_{0})
    \dif{\psi_{0}}
    =
    \frac{2A(v,\alpha )}{\sqrt{r^{2}-b^{2}}}
    \end{equation}
%
for $ r> b $ and $\overline {f} (v,\alpha; r) = 0 $ for $ r <b $. Having integrated the function $\overline {f} $ over the radius $ r $, we can express the normalization constant $ A (v,\alpha) $ in terms of the total number of ions $ N (v,\alpha) $ with the given speed $ v $ and incidence angle $\alpha $:
    \begin{equation}
    \label{2:04}
    N(v,\alpha)
    =
    \int_{b}^{a} 2\pi\, r\, \overline{f}(v,\alpha ; r) \dif{r}
    =
    4\pi A(v,\alpha )\, \sqrt{a^{2}-b^{2}}
    .
    \end{equation}
Consequently,
    \begin{multline}
    \label{2:05}
    \overline{f}(v,\alpha; r) = \frac{N(v,\alpha)}{
        2\pi
        \sqrt{a^{2}-b^{2}}
        \sqrt{r^{2}-b^{2}}
    }
    =
    \\
    =
    \frac{N(v,\alpha)}{
        2\pi a \left|\cos\alpha \right|
        \sqrt{r^{2}-a^{2}\sin^{2}\alpha}
    }
    \end{multline}
%
for $ a\sin\alpha <r <a $.
By integrating over the angle $\alpha $ and the velocity $ v $, we find the radial ion density profile inside the cylinder:
    \begin{equation}
    \label{2:06}
    n(r)
    =
    2
    \int_{0}^{\infty } \dif{v} v
    \int_{-\arcsin(r/a)}^{+\arcsin(r/a)}
    \overline{f}(v,\alpha; r)\dif\alpha
    .
    \end{equation}
%
The reason for adding coefficient 2 to this formula will be explained later. A uniform density distribution inside the cylinder $ n (r) = n $ is obtained if
    \begin{equation}
    \label{2:07}
    N(v,\alpha ) = f(v)\, a^{2}\cos^{2}\alpha
    ,
    \qquad
    2\pi\int_{0}^{\infty }f(v)\,v\dif{v} = n
    .
    \end{equation}
Then
    \begin{equation}
    \label{2:08}
    \overline{f}(v,\alpha; r)
    =
    \frac{ a\cos\alpha}{
        \sqrt{r^{2}-a^{2}\sin^{2}\alpha}
    }\,f(v)
    .
    \end{equation}
%
It may seem surprising, but the function \eqref{2:08} describes the isotropic (along the angle $ \theta $) distribution of ions. To prove this, we reproduce the derivation of Eq.~\eqref{2:08} in another way.

%
Let $ f (v, \theta; r, \psi) $ be the distribution function of ions in phase space, so that
    \begin{equation}
    \label{2:11}
    n(r,\psi)
    =
    \int_{0}^{\infty } \dif{v} \int_{-\pi}^{\pi} \dif\theta\,
        v\,f(v,\theta; r,\psi)
    .
    \end{equation}
%
In an inhomogeneous plasma, the function $ f (v, \theta; r, \psi) $ can actually depend on the spatial coordinates of $ (r, \psi) $, but for brevity we do not write them further in the arguments of the distribution function $ f $ and its moments like density $ n $, current density $ \vec {j} $, etc. Turning to Fig.~\ref{fig:CoordinatesInCylinder}, we compose the equations
    \begin{equation}
    \label{2:12}
    \begin{gathered}
    r^2=a^2-2 a q \cos (\alpha )+q^2
    ,\\
    a^2=q^2+2 q r \cos (\theta -\psi )+r^2
    .
    \end{gathered}
    \end{equation}
%
The first equation expresses the length $ r $ of one of the sides of the OAB triangle through the opposite angle $ \alpha $ and the lengths of $ a $ and $ q $ of the other two sides. The second equation connects the length $ a $ with the opposite angle $ \pi- \theta + \psi $. Eliminating $ q $ from these equations yields Eq.~\eqref{2:02a}.
%
It also follows from the equality of the common leg OC in the triangles OAC and OBC. Now we can calculate the Jacobian of the transform
    \begin{equation}
    \label{2:15}
    \parder{(v,\theta)}{(v,\alpha) }
    =
    \parder{\theta }{\alpha }
    =
    \frac{a\cos(\alpha )}{\sqrt{r^{2}-a^{2}\sin^{2}(\alpha )}}
    ,
    \end{equation}
in order to replace the variables $ (v, \theta) $ in the integral \eqref{2:11} with a pair of variables $ (v, \alpha) $. As a result, we get
    \begin{equation}
    \label{2:16}
    n = 2\int_{0}^{\infty } v\dif{v} \int_{-\arcsin(r/a)}^{+\arcsin(r/a)} \dif\alpha \overline{f}(v,\alpha)
    ,
    \end{equation}
where
    \begin{equation}
    \label{2:17}
    \overline{f}(v,\alpha)
    =
    \parder{\theta }{\alpha }
    f(v,\theta)
    =
    \frac{a\cos(\alpha )}{\sqrt{r^{2}-a^{2}\sin^{2}(\alpha )}}
    f(v,\theta(\alpha))
    ,
    \end{equation}
%
and the coefficient $ 2 $ is due to the fact that the angle $ \alpha $ twice passes the interval $ - \arcsin (r/a) <\alpha <\arcsin (r/a) $ when $ \theta $ changes from $ - \pi $ up to $ + \pi $. With an isotropic ion distribution, the function $ f (v, \theta) = f (v) $ does not depend on the angle $ \theta $, and the function \eqref{2:17} coincides with the function \eqref{2:08}, which completes the proof.

%
Here are some examples. Let there be only ions with one single value of the angle of incidence inside the diamagnetic bubble. Substituting
    \begin{equation}
    \label{2:21}
    f(v,\alpha ) = \frac{n_{}}{V}\,\delta(v-V)\,\delta(\sin(\alpha )-b/a)
    \end{equation}
in Eqs.~\eqref{2:16} and~\eqref{2:17}, yields
    \begin{equation}
    \label{2:22}
    n(r) = \frac{an_{0}}{\sqrt{r^{2}-b^{2}}}
    \end{equation}
%
for $ r> b $ and $ n (r) = 0 $ for $ r <b $. Such a density distribution has an integrable singularity at $ r \to b $.

%
As a second example, we consider the situation when ions with a small angle of incidence die on the diaphragm, because they penetrate too deeply into the boundary layer. Such a distribution is modeled by the distribution function
    \begin{equation}
    \label{2:23}
    f(v,\alpha ) = \frac{2n_{0}}{\pi V}\,\delta(v-V)\,\Heviside(\sin(\alpha )-b/a),
    \end{equation}
where $ \Heviside $ denotes the Heaviside function such that $ \Heviside (x) = 1 $ for $ x \ge0 $ and $ \Heviside (x) = 0 $ for $ x <0 $. By integrating such a distribution function, a density profile is obtained in the form of a ring in which the density monotonically increases from zero at $ r = b $ to the maximum value at $ r = a $:.
    \begin{equation}
    \label{2:24}
    n(r) = n_{0} \left[
        1
        -
        \frac{2}{\pi}\arctan\frac{b}{\sqrt{r^{2}-b^{2}}}
    \right]
    .
    \end{equation}

%
Finally, we note that in the case of an isotropic distribution of ions inside the diamagnetic bubble, when the distribution function in the phase space does not depend on $ \theta $, i.e., $ f (v, \theta) = f (v) $, and the density and the pressures are homogeneous, the distribution function $ \overline {f} $ in the variables $ (v, \alpha) $ does not depend on the angle of incidence $ \alpha $ at the inner boundary of the bubble. Indeed, for $ r = a $ from Eq.~\eqref{2:02a} we find that $ \alpha = \theta $, and from Eq.~\eqref{2:17} it follows that
    \begin{equation}
    \label{2:25}
    \overline{f}(v,\alpha)
    =
    f(v)
    .
    \end{equation}
In the next section, we will consider this particular case.

\section{Flat Boundary Layer}\label{s3}

\begin{figure}
  \centering
  \includegraphics[width=\linewidth]{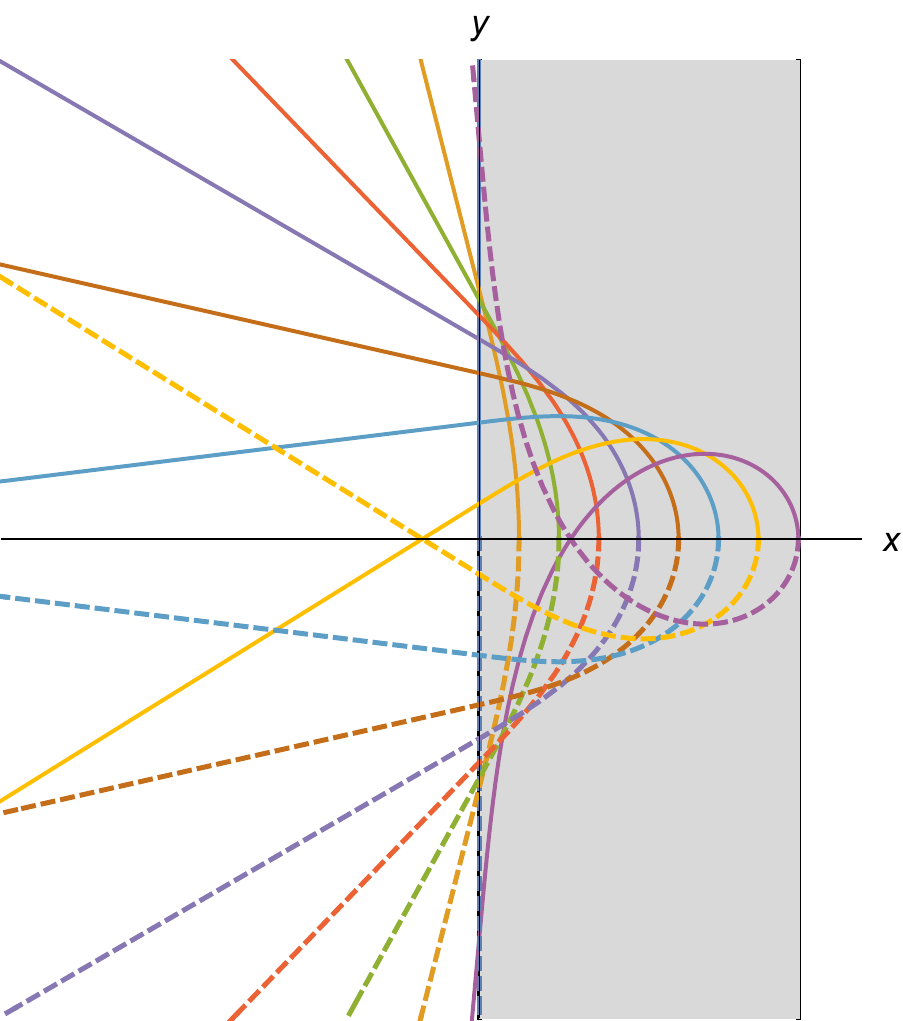}
  \caption{
    %
    The trajectories of the ions in a flat boundary layer. The boundary layer is shown in gray. The magnetic field is zero to the left of the layer, increases linearly inside the layer from zero to the maximum value, which is equal to the field to the right of the layer. Incoming and outgoing branches of the trajectories are depicted by solid and dashed lines, respectively. The angle $ \alpha $, which determines the direction of the velocity according to Eq.~\eqref{3:01}, is counted from the direction of the $ x $ axis. The paths that enter from above correspond to the negative angle of incidence $ \alpha_{0} <0 $. The trajectories with $ \alpha_{0}> 0 $, which enter from below, penetrate deeper into the boundary layer.
  }\label{fig:MotionInPlanarLayer}
\end{figure}

%
Suppose that a diamagnetic bubble has such a large radius that with sufficient accuracy a small portion of the boundary layer can be considered flat, as shown in Fig.~\ref{fig:MotionInPlanarLayer}. We direct the axis $ x $ deep into the layer along the normal and introduce the angle $ \alpha $ between the direction of speed and the axis $ x $, so that
    \begin{equation}
    \label{3:01}
    \vec{V} = \{V\cos\alpha , V\sin\alpha , V_{z}\}
    .
    \end{equation}
For simplicity, we first consider the case when all ions have the same kinetic energy and, accordingly, the same velocity modulus $ V $. The magnetic field is directed along the $ z $ axis perpendicular to the plane of the figure and depends on the normal coordinate $x$, namely
    \begin{equation}
    \label{3:02}
    \vec{B} = \{0, 0, B(x)\}
    .
    \end{equation}
Projecting the equation of motion
    \begin{equation}
    \label{3:03}
    \dot{\vec{V}} =
    (e/mc)
    \left[
        \vec{V}\times \vec{B}
    \right]
    \end{equation}
on the $ x $ axis, we obtain the equation
    \begin{equation}
    \label{3:04}
    V\cos(\alpha(x))\,\alpha'(x) = - \Omega (x)
    ,
    \end{equation}
in which $ \Omega (x) = eB (x)/mc $ denotes the cyclotron frequency of the ions, and the prime denotes the derivative with respect to the coordinate $ x $. Integrating it, we obtain the equation
    \begin{equation}
    \label{3:05}
    \sin(\alpha)
    =
    \sin(\alpha_{0})
    -
    \frac{1}{V}\int_{0}^{x} \Omega(x)\dif{x}
    ,
    \end{equation}
%
that relates the angle of incidence $ \alpha_{0} $ of the ion on the inner side of the boundary layer at $ x = 0 $ with the penetration depth of the particle inside the layer. It follows from this that $ \sin(\alpha) $ decreases as the ion moves deeper into the boundary layer. The maximum depth $ \delta (\alpha_{0}) $ of penetration into the boundary layer corresponds to $ \sin \alpha = -1 $. For a known magnetic field profile inside the boundary layer, the value $ \delta (\alpha_{0}) $ can be found from the equation
    \begin{equation}
    \label{3:06}
    1+\sin(\alpha_{0})
    =
    \frac{1}{V}\int_{0}^{\delta} \Omega(x)\dif{x}
    .
    \end{equation}
%
If $ \alpha_{0} = - \pi/2 $, then we have a case of sliding incidence, and $ \delta = 0 $, that is, the depth of penetration of the particle into the boundary layer is zero. The particle hardly scatters and
    \begin{equation}
    \label{3:07}
    \frac{1}{V}\int_{0}^{\delta} \Omega(\xi)\dif\xi
    \to 0.
    \end{equation}
%
If $ \alpha_{0} = 0 $, then we have a case of normal incidence. In this case
    \begin{equation}
    \label{3:08}
    \frac{1}{V}\int_{0}^{\delta} \Omega(\xi)\dif\xi
    \to 1
    ,
    \end{equation}
%
and the particle is scattered by 180 degrees (reflected from the boundary along the normal). If $ \alpha_{0} = + \pi/2 $, then the particle almost does not scatter again, but flies to the boundary from the opposite side. In this case, the maximum penetration depth for a given speed $ V $ is achieved so that
    \begin{equation}
    \label{3:09}
    \frac{1}{V}\int_{0}^{\delta} \Omega(\xi)\dif\xi
    \to 2.
    \end{equation}

%
To simplify the following reasoning, let us assume for a while that on the boundary of a plane layer with $ x = 0 $ all ions have the same (in absolute value) velocity $ V $, and their distribution over the angle of incidence $ \alpha_{0} $ is isotropic. Formally, this assumption corresponds to the distribution function
    \begin{equation}
    \label{3:11}
    \overline{f}(v,\alpha_{0})
    =
    \frac{n_{0}}{2\pi V}\,\delta (v-V)
    .
    \end{equation}
%
Later, we will get rid of the assumption that the velocity distribution is monochromatic, but now it will allow us not to clutter the formulas with an additional integral over $ v $ and operate with the density $ n $ and the velocity $ V $ instead of the integrals.

%
First of all, we calculate the density $ n $ inside the boundary layer. This is easy to do, noting that in a stationary system the particle flux normal to the layer on the incoming branch of the ion path with a given angle of incidence $ \alpha_{0} $ is a conserved quantity, i.e.
    \begin{equation}
    \label{3:12}
    \dif j_{x} = e \dif{n} V_{x}
    = \const
    .
    \end{equation}
Since $ V_{x} = V \cos (\alpha) $, from the constancy of $ \dif{j_{x}} $ we immediately find
    \begin{equation}
    \label{3:14}
    \dif{n}
    =
    \frac{n_{0}}{2\pi}\, \frac{\cos(\alpha_{0})}{\cos(\alpha )}\dif{\alpha_{0}}
    ,
    \end{equation}
and besides
    \begin{equation}
    \label{3:15}
    \cos(\alpha ) = \sqrt{
        1 - \left(
            \sin(\alpha_{0}) - G/V
        \right)^{2}
    }
    ,
    \end{equation}
where
    \begin{equation}
    \label{3:16}
    G = (e/mc) \int_{0}^{x} B(x)\dif{x}
    .
    \end{equation}
%
As shown in Fig.~\ref{fig:MotionInPlanarLayer}, the incoming and outgoing branches of the ion paths are completely symmetrical, therefore the formula \eqref{3:14} is true for the entire boundary layer. Now, to calculate the density at a certain depth of the layer $ x $, we need to integrate over the angle of incidence $ \alpha_{0} $ and multiply the integral by $ 2 $ to take into account both branches of the trajectory. Taking into account that ions with the incidence angle $ \alpha_{0} $ in the interval from $ \arcsin (-1 + G/V) $ to $ + \pi/2$ penetrate into the boundary layer at a distance with a given value of $ G$, we get
    \begin{multline}
    \label{3:17}
    n(x)
    =
    \frac{n_{0}}{2\pi}
    2 \int_{\arcsin(-1+G/V)}^{\pi/2}
    \frac{
        \cos(\alpha_{0})
    }{
        \cos(\alpha)
    }
    \dif{\alpha_{0}}
    \\
    =
    \frac{n_{0}}{\pi}
    \left(
        \pi/2
        +
        \arcsin(1-G(x)/V)
    \right)
    .
    \end{multline}
Because
    \begin{equation}
    \label{3:18}
    \dif J_{y} = e \dif n V_{y} = e \dif n V \sin(\alpha )
    ,
    \end{equation}
we need to add the factor $ \sin (\alpha) $ in the integrand to calculate the tangential component of the ion current density in the boundary layer:
    \begin{multline}
    \label{3:19}
    J_{y}(x)
    =
    \frac{en_{0}}{2\pi}
    2 \int_{\arcsin(-1+G/V)}^{\pi/2}
    \frac{
        \cos(\alpha_{0})
    }{
        \cos(\alpha)
    }
    \sin(\alpha)
    \dif{\alpha_{0}}
    \\
    =
    -
    \frac{en_{0}}{\pi}
    \sqrt{G\left(2V-G\right)}
    .
    \end{multline}
%
The normal component of the current density $ j_{x} $ is equal to zero, since the contributions of the incoming and outgoing branches of the trajectories cancel each other out. For $ G> 2V $, $ n $ and $ J_{y} $ also vanish. Similarly, we calculate the components of the momentum flux tensor:
    \begin{multline}
    \label{3:21}
    \Pi_{xx}
    =
    \frac{m n_{0}}{2\pi}
    2 \int_{\arcsin(-1+G/V)}^{\pi/2}
    \frac{
        \cos(\alpha_{0})
    }{
        \cos(\alpha)
    }
    \left[
        V\cos(\alpha)
    \right]^{2}
    \dif{\alpha_{0}}
    =
    \\
    =
    \frac{m n_{0}}{2\pi}
    \left[
        V^2\arccos\left(G/V-1\right)
        +
        \sqrt{(2 V-G)\,G}\, (V-G)
    \right]
    ,
    \end{multline}
    \begin{multline}
    \label{3:22}
    \Pi_{yy}
    =
    \frac{m n_{0}}{2\pi}
    2 \int_{\arcsin(-1+G/V)}^{\pi/2}
    \frac{
        \cos(\alpha_{0})
    }{
        \cos(\alpha)
    }
    \left[
        V\sin(\alpha)
    \right]^{2}
    \dif{\alpha_{0}}
    =
    \\
    =
    \frac{m n_{0}}{2\pi}
    \left[
        V^2\arccos\left({G}/{V}-1\right)
        -\sqrt{(2 V-G)\,G}\, (V-G)
    \right]
    .
    \end{multline}
They are related to the components of the pressure tensor by the relations
    \begin{gather}
    \label{3:23}
    P_{xx} = \Pi_{xx}
    ,\qquad
    P_{yy} = \Pi_{yy} - \frac{1}{2} m n U_{y}^{2}
    ,
    \end{gather}
where
    \begin{equation}
    \label{3:24}
    U_{y} = \frac{J_{y}}{en}
    =
    \frac{
        \sqrt{G\left(2V-G\right)}
    }{
        \pi/2 + \arcsin(1-G/V)
     }
    .
    \end{equation}
On the inner side of the boundary layer with $ G = 0 $ we have
    \begin{equation}
    \label{3:25}
    P_{xx} = P_{yy} = \frac{1}{2} n_{0}mV^{2}
    ,
    \end{equation}
on the outside, when $ G = 2V $, the pressure vanishes, $ P_{xx} = P_{yy} = 0 $. Inside the boundary layer $ P_{xx}> P_{yy} $.

\section{Magnetic field in flat layer}\label{s4}

To find the magnetic field profile in the boundary layer, one needs to solve the equation
    \begin{equation}
    \label{4:01}
    B'(x) = - \frac{4\pi}{c} J_{y}
    .
    \end{equation}
Taking Eqs.~\eqref{3:16} and~\eqref{3:19} into account, we obtain a second-order ordinary differential equation for the function $G(x)$,
    \begin{equation}
    \label{4:02}
    G''(x)
    =
    -
    \frac{4 e^{2} n_{0}}{mc^{2}}\,
    \sqrt{G\left(2V-G\right)}
    ,
    \end{equation}
which must be supplemented by the boundary conditions $G = 0$, $G'= 0$ for $ x = 0 $. The solution can be expressed in quadratures as an implicit function $ x = x (G) $. In the limit $ x \to0 $ it is reduced to the power functions
    \begin{gather}
    \label{4:03}
    G = \frac{2e^{4}n_{0}^{2}V}{9m^{2}c^{4}}\,x^{4}
    =
    \frac{2eB_{\textnormal{v}}\rho_{i}}{9\pi^{2}mc}\left(
        \frac{x}{\rho_{i}}
    \right)^{4}
    ,\\
    \label{4:04}
    B = \frac{8e^{3}n_{0}^{2}V}{9mc^{3}}\,x^{3}
    =
    \frac{B_{\textnormal{v}}}{18\pi^{2}}\left(
        \frac{x}{\rho_{i}}
    \right)^{3}
    ,
    \end{gather}
where the notation
    \begin{equation}
    \label{4:05}
    B_{\textnormal{v}} = \sqrt{4\pi m n_{0}V^{2}}
    ,
    \end{equation}
is introduced for the vacuum magnetic field beyond the boundary layer,
    \begin{equation}
    \label{4:06}
    \rho_{i}
    =
    \frac{mcV}{eB_{\textnormal{v}}}
    =
    \frac{c}{\omega_{pi}}
    \end{equation}
denotes the Larmor radius of ions in the vacuum magnetic field, and
    \begin{equation}
    \omega_{pi} = \sqrt{4\pi e^{2}n_{0}/m}
    \end{equation}
makes the sense of the plasma ion frequency. Multiplying Eq.~\eqref{4:01} by $ B = (mc/ e) \, G'$ and performing integration over $ x $, we can verify that the result is identical to the transverse equilibrium condition
    \begin{equation}
    \label{4:07}
    \frac{B^{2}}{8\pi} + P_{xx}
    =
    \frac{1}{2} mn_{0}V^{2}
    =
    \frac{B_{\textnormal{v}}^{2}}{8\pi}
    .
    \end{equation}
Consequently, on the outside of the boundary layer and behind it
    \(
    B=B_{\textnormal{v}}
    \),
therefore, $ B_{\textnormal{v}} $ really makes sense of a vacuum magnetic field. Inside the layer near its outer side at $ x <x_ \textnormal{out} $
    \begin{equation}
    \label{4:09}
    B
    =
    B_{\textnormal{v}}
    -
    \frac{2^{3/2}}{3\pi}
    \left(
        \frac{x_\textnormal{out}-x}{\rho_{i}}
    \right)^{3/2}
    B_{\textnormal{v}}
    .
    \end{equation}

\begin{figure*}
  \centering
  \includegraphics[width=0.47\linewidth]{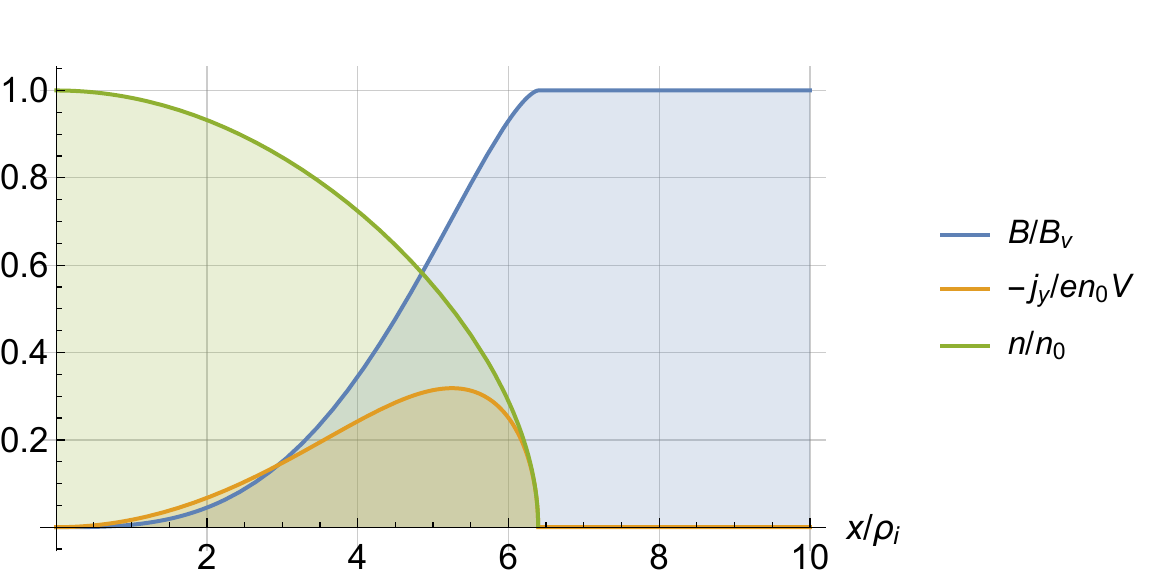}\hfil
  \includegraphics[width=0.47\linewidth]{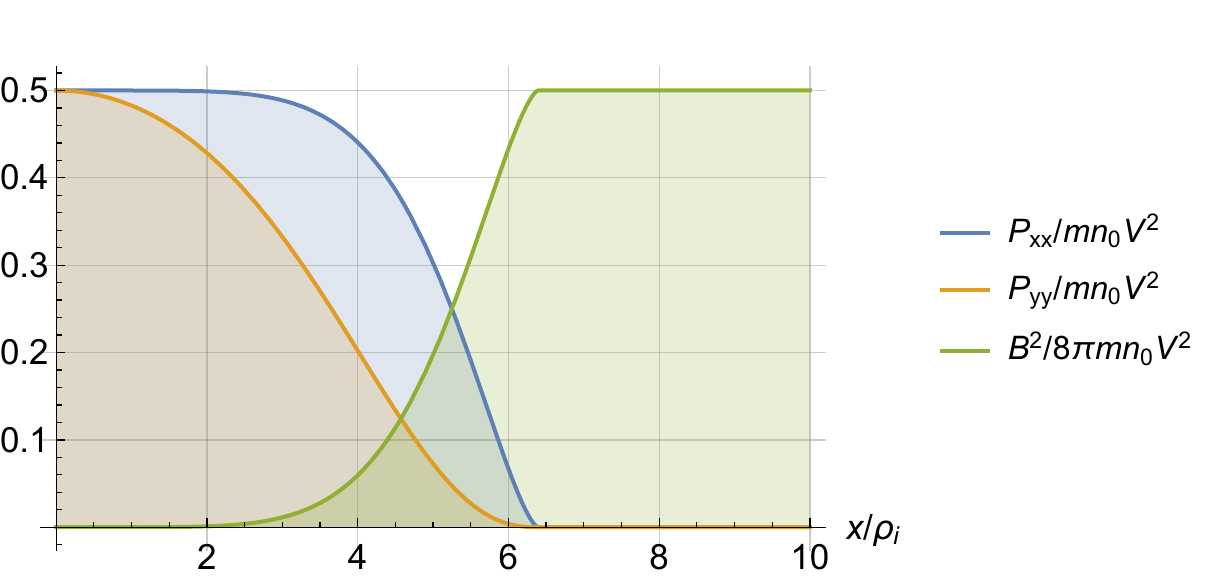}
  \caption{
    Profiles of the magnetic field, current density and ion density (left) and profiles of transverse gas-kinetic pressure and magnetic field pressure (right) in a flat boundary layer with a monochromatic ion energy distribution.
  }
  \label{fig:B-J-n(Mono)}
\end{figure*}
%
Profiles of the main parameters in the boundary layer are shown in Fig.~\ref{fig:B-J-n(Mono)} depending on the coordinate $ x $ normalized by the Larmor radius $\rho_{i}$. As can be seen from the figure, the thickness of the boundary layer exceeds six Larmor radii (the exact value is $ x_\textnormal{out} = \num {6.39499} \rho_{i} $).


The resulting solution has a logical flaw. It does not explain why suddenly at some point $ x = 0 $ the magnetic field begins to grow from a zero value.

\section{Distribution function in a flat layer}\label{s5}

%
A stationary solution of the collisionless kinetic equation can be an arbitrary function $ F $ of the integrals of motion. In axially symmetric diamagnetic bubble, such integrals of motion are the kinetic energy $ \varepsilon = \frac {1} {2} mv^{2} $ and the azimuthal component of the generalized angular momentum $ p_{\psi} = mv_{ \psi} r + (e/c) \, rA_{\psi} $, where $ A_{\psi} (r,z) $ denotes the corresponding component of the vector potential. An example is the function
    \begin{equation}
    \label{5:01}
    F(\varepsilon,p_{\psi})
    =
    \frac{m n_{0}}{2\pi T}\,
    \exp\left[
        -\frac{mv^{2}-2\omega p_{\psi}}{2T}
    \right]
    ,
    \end{equation}
%
that describes the solid-state rotation of particles around the axis of a cylinder with a frequency $ \omega $.


In the one-dimensional system studied in Sections \ref{s3} and \ref{s4}, conserving quantity is the $y$-component of the generalized momentum $ p_{y} = mv_{y} + \left (e/c \right) A_{y } = m \left (v_ {y} + G \right) $. Therefore, it is necessary to find such a function $ f (v, \theta) = F (\varepsilon, p_{y}) $ that would coincide with the isotropic function $ f (v) $ inside the diamagnetic bubble, where $ G = 0 $ and $ p_{y} = mv_{y} $.

%
Let us again turn to the imaginary case when only particles with one single energy value are present in the distribution of ions, i.e.,\
    \begin{equation}
    \label{5:02}
    f(v) = \frac{n_{0}}{2\pi V}\, \delta (v-V)
    .
    \end{equation}
%
We will get rid of this assumption at the end of this section. The generalized momentum $ p_{y} $ inside the bubble varies from $ mV $ at the inner boundary of the surface layer, where $ G = 0 $, to $ - \infty $ inside the bubble, given that $ G \leq0 $ there, and in this interval, the distribution function $ F $ should not depend on $ p_{y} $, otherwise the isotropy would be violated, and for $ p_{y}> mV $ there should be $ F = 0 $, since there are no particles with such $ p_ {y } $ in the distribution. Thus,
    \begin{equation}
    \label{5:03}
    f(v,\theta )
    =
    \frac{n_{0}}{2\pi V}\,\delta(v-V)\,
    \Heviside(mv-p_{y})\,
    ,
    \end{equation}
%
where $\Heviside$ denotes the Heaviside function, which is equal to one if its argument is non-negative, and equal to zero if it is less than zero. Note that in the limit of a small Larmor radius, the formula \eqref{5:03} goes into
    \begin{equation}
    \label{5:03a}
    f(v,\theta )
    =
    \frac{n_{0}}{2\pi V}\,\delta (v-V)\,
    \Heviside(-(e/c)\,A_{y}(X)\,)\,
    ,
    \end{equation}
%
where $ X $ is the coordinate of the leading center of the Larmor orbit of the ion. Obviously, in this case it describes the plasma density and pressure profile in the form of a step. Recall also that the function $ f (v, \theta) $ defines the particle density in the phase space $ (v, \theta) $, in contrast to the distribution function $ \overline {f} (v, \alpha) = (\tparder{\theta}{\alpha}) \, f (v, \theta) $, which has the meaning of density in the configuration space of the variables $ (v $, $ \alpha) $.
%
%
Substituting $ v_{y} = v \sin (\theta) $ in Eq.~\eqref{5:03}, we find the final expression for the distribution function
    \begin{equation}
    \label{5:04}
    f(v,\theta)
    =
    \frac{n_{0}}{2\pi V}\,\delta (v-V)\,
    \Heviside\!\left(
        v
        -
        \left(v\sin(\theta)+G\right)
    \right)
    .
    \end{equation}
%
Substitution of this function into the integral \eqref{2:11} gives the expression \eqref{3:17} already found earlier for the ion density in the boundary layer. Similarly, the current density \eqref{3:19} and the components of the momentum flux tensor \eqref{3:21} and \eqref{3:22} can be recalculated.

\begin{figure*}
  \centering
  \includegraphics[width=0.47\linewidth]{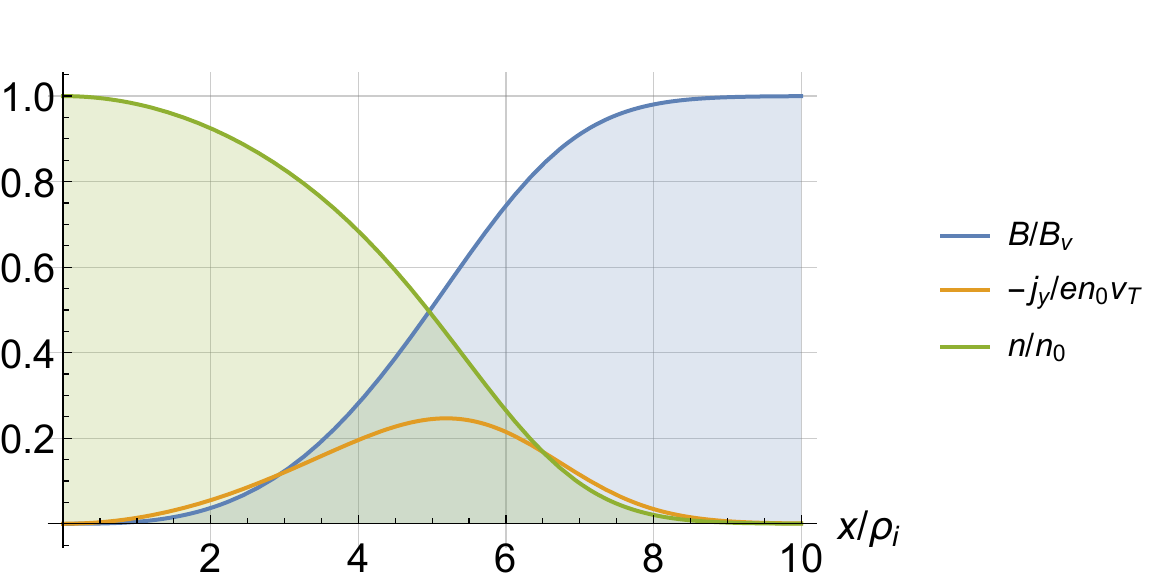}\hfil
  \includegraphics[width=0.47\linewidth]{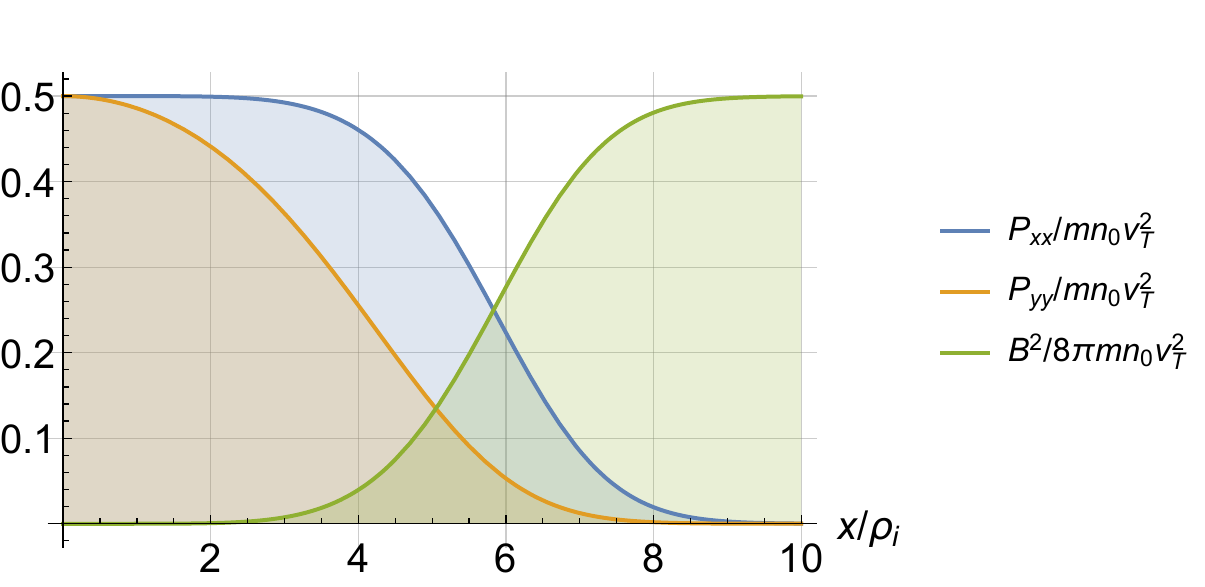}
  \caption{
    Profiles of the magnetic field, current density and ion density (left) and profiles of transverse gas-kinetic pressure and magnetic field pressure (right) in a flat boundary layer with a Maxwellian energy distribution of ions.
  }
  \label{fig:B-J-n(Maxwell)}
\end{figure*}

%
By adjusting Eq.~\eqref{5:04}, we can simulate an arbitrary distribution of ions over energies. In particular, if ions inside the diamagnetic bubble have a Maxwell distribution with a thermal velocity $ v_{T} $, then
    \begin{equation}
    \label{5:5}
    f(v,\theta)
    =
    \frac{n_{0}}{2\pi v_{T}^{2}}\,
    \exp\left(-\frac{v^{2}}{v_{T}^{2}}\right)
    \Heviside\!\left(
        v
        -
        \left(v\sin(\theta)+G\right)
    \right)
    .
    \end{equation}
%
For this case, it is possible to calculate the current density $ J_{y} $ in an analytical form, but other quantities can only be calculated numerically. The calculation results are shown in Fig.~\ref{fig:B-J-n(Maxwell)}. As can be seen from the figure, the outer boundary of the transition layer is blurred and formally extends to infinity. It can be approximately assumed that the width of the boundary layer is 8 Larmor radii $ \rho_{i} = mcv_{T}/eB_ {\textnormal{v}} $ calculated from the thermal velocity $ v_{T} $ in a vacuum magnetic field $ B_{\textnormal{v}} = \sqrt {4 \pi n_{0} mv_ {T}^{2}} $. Note that $ \rho_{i} = c/\omega_{pi} $, where $ \omega_ {pi} = \sqrt {4 \pi n_{0} e^{2}/m} $ denotes the plasma ion frequency.

\section{Boundary layer in a cylinder}\label{s6}

We introduce the magnetic flux function
    \begin{equation}
    \label{6:01}
    \chi(r) = r A_{\psi}(r)
    ,
    \end{equation}
such that
    \begin{equation}
    \label{6:02}
    B(r) = \frac{1}{r}\,\chi'(r)
    ,
    \end{equation}
and suppose that it is equal to zero on the axis of the system at $ r = 0 $ so that
    \begin{equation}
    \label{6:03}
    \chi(r) = \int_{0}^{r} r\,B(r)\dif{r}
    .
    \end{equation}
We emphasize that in the current Section the dependence of the parameters of the diamagnetic bubble on the coordinate $ z $ is ignored. We again assume that all ions have the same energy $ \varepsilon = mV^{2}/2 $, and their distribution over the azimuthal angle $ \theta $ is isotropic inside the diamagnetic bubble. Turning to Fig.~\ref{fig:CoordinatesInCylinder}, the generalized azimuthal moment can be written as
    \begin{equation}
    \label{6:03a}
    p_{\psi}
    =
    mrv\sin(\theta -\psi) + (e/c)\,\chi(r)
    ,
    \end{equation}
where $ v $ is the projection of velocity onto the plane $ z = \const $; Ion motion along the $ z $ axis is ignored for now. Taking into account that $ \varepsilon $ and $ p_{\psi} $ are integrals of motion, by analogy with Eq.~\eqref{5:03} we write the ion distribution function in the form
    \begin{multline}
    \label{6:04}
    f(v,\theta)
    =
    \frac{n_{0}}{2\pi V}\,\delta (v-V)\,
    \times
    \\
    \Heviside\!\left(
        av + (e/mc)\,\chi(a)
        -
        \left[
            r v\sin(\theta-\psi) + (e/mc)\,\chi(r)
        \right]
    \right)
    .
    \end{multline}
The result of calculating various moments of this distribution function is written below using the notation
    \begin{equation}
    \label{6:05}
    G = \left(e/mc\right)
    \left( \chi(r) - \chi(a) \right)/V
    ,
    \end{equation}
with the dimension of length. The integration of the Heaviside function, which is included in Eq.~\eqref{6:04}, is not difficult, but the result is cumbersome due to the abundance of options in the size of the integration region over the angle $ \theta $ for different combinations of the values $ r $, $ a $ and $ G $.
\begin{widetext}
We give the formulas for the density of ions
    \begin{equation}
    \label{6:06}
    n
    =
    \int_{0}^{\infty } \dif{v} \int_{-\pi}^{\pi} \dif\theta v\,f(v,\theta )
    =
    \frac{n_{0}}{\pi }
    \begin{cases}
    0
        & G\geq a+r 
    \\
    \arcsin\left(\frac{a-G}{r}\right)
    +
    \frac{\pi}{2}
        & 
        a-r < G < a+r
    \\
    \pi
        & \textnormal{True}
    \end{cases}
    \end{equation}
and azimuthal components of current density
    \begin{equation}
    \label{6:07}
    J_{\psi}
    =
    \int_{0}^{\infty } \dif{v} \int_{-\pi}^{\pi} \dif\theta
    v^{2}\sin(\theta-\psi)\,f(v,\theta )
    =
    -
    \frac{en_{0}V}{\pi r}
    \begin{cases}
    0
        & G\geq a+r 
    \\
    \sqrt{r^{2} - (a-G)^{2}}
        & 
        a-r < G < a+r
    \\
    0
        & \textnormal{True}
    \end{cases}
    .
    \end{equation}
A consequence of axial symmetry is that $ n $ and $ J_{\psi} $ are independent of the azimuthal angle $ \psi $. We will also find useful expressions for the components of the momentum flux tensor:
    \begin{gather}
    \label{6:08}
    \Pi_{rr}
    =
    m\int_{0}^{\infty } \dif{v} \int_{-\pi}^{\pi} \dif\theta v^{3}\cos^{2}(\theta-\psi)\,f(v,\theta )
    =
    \frac{mn_{0}V^{2}}{2\pi }
    \begin{cases}
    0
        & G\geq a+r 
    \\
    \frac{\sqrt{r^2-(a-G)^2} (a-G)}{r^2}
    +
    \arcsin\left(\frac{a-G}{r}\right)
    +
    \frac{\pi}{2}
        & 
        a-r < G < a+r
    \\
    \pi
        & \textnormal{True}
    \end{cases}
    ,
    \\
    \label{6:09}
    \Pi_{\psi \psi }
    =
    m\int_{0}^{\infty } \dif{v} \int_{-\pi}^{\pi} \dif\theta v^{3}\sin^{2}(\theta-\psi)\,f(v,\theta )
    =
    \frac{mn_{0}V^{2}}{2\pi }
    \begin{cases}
    0
        & G\geq a+r 
    \\
    -\frac{\sqrt{r^2-(a-G)^2} (a-G)}{r^2}
    +
    \arcsin\left(\frac{a-G}{r}\right)
    +
    \frac{\pi}{2}
        & 
        a-r < G < a+r
    \\
    \pi
        & \textnormal{True}
    \end{cases}
    .
    \end{gather}
\end{widetext}
Later in this section, we omit the dimensional factors $ n_{0} $, $ en_ {0} V $ and $ mn_{0} V^{2} $ in the intermediate formulas.


The magnetic flux \eqref{6:03} is found as a solution to the ordinary differential equation
    \begin{equation}
    \label{6:11}
    \frac{1}{r}\chi'' - \frac{1}{r^{2}} \chi'
    =
    - \frac{4\pi}{c} J_{\psi}
    \end{equation}
with boundary conditions
    \begin{equation}
    \label{6:12}
    \chi=0,
    \qquad
    \chi'/r=B_{\textnormal{in}}
    \end{equation}
at $r=0$.\footnote{
    In fact, we searched for the function $ \overline \chi (r) = \chi (r) - \chi (a) $ and set the boundary conditions $ \overline \chi = 0 $ and $ \overline {\chi} '/ r = B_{\textnormal{in}} $ for $ r = a $. By calculating $ \overline \chi (r) $ we found $ \chi (r) = \overline \chi (r) - \overline \chi (0) $.
}

\begin{figure*}
  \centering
\includegraphics[width=0.47\linewidth]{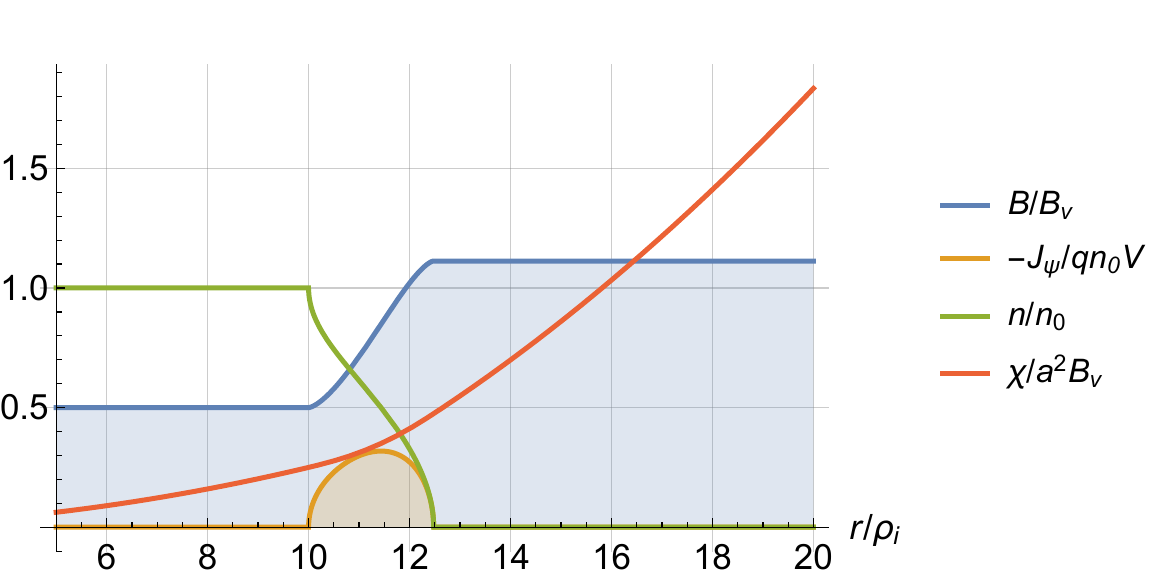}\hfil
\includegraphics[width=0.47\linewidth]{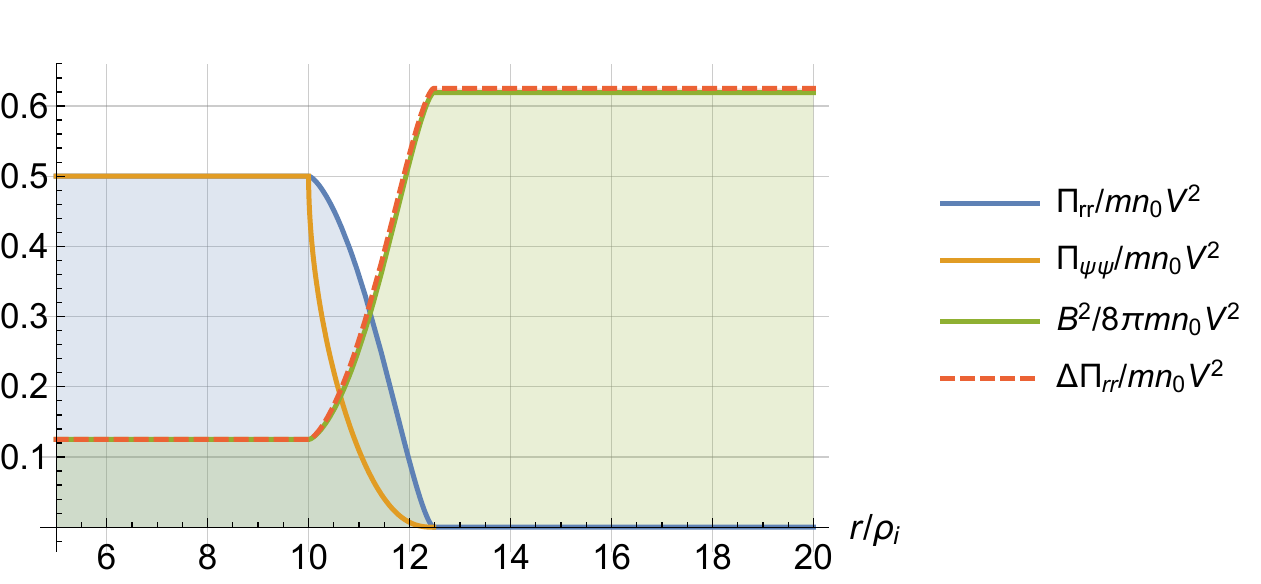}\hfil
\caption{
    Radial profiles of the magnetic field, current density, density (left) and components of the momentum flux tensor, magnetic field pressure (right) for $ a/\rho_{i} = 10 $ and $ B_{\textnormal{in}}/B_{\textnormal{v}} = 5 \rho_{i}/a $. The dashed line shows the graph of the function $ B^{2}/8 \pi + \Delta \Pi_{rr} $.
  }\label{fig:(Bin,0_5)}
\end{figure*}


\begin{figure*}
  \centering
\includegraphics[width=0.47\linewidth]{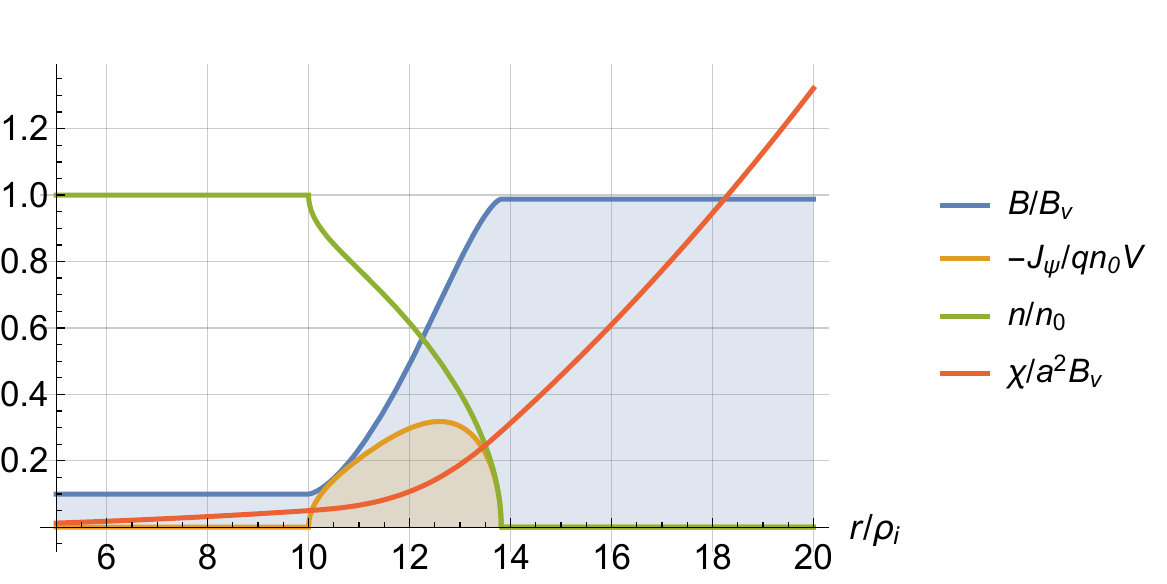}\hfil
\includegraphics[width=0.47\linewidth]{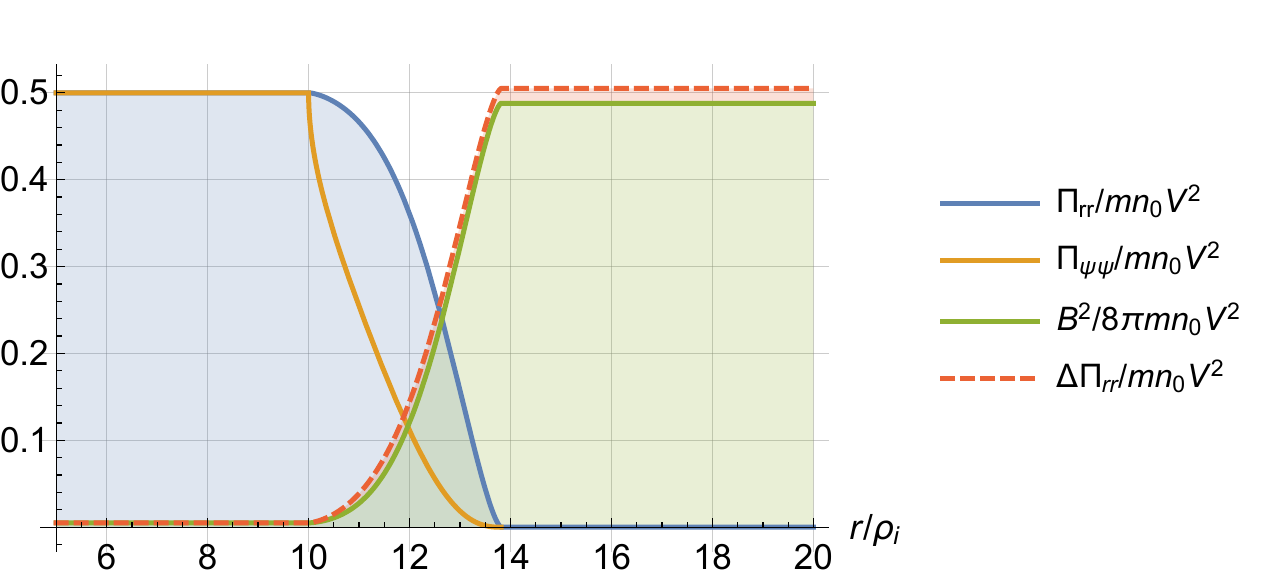}
\caption{
    Radial profiles of the magnetic field, current density, density (left) and components of the momentum flux tensor, magnetic field pressure (right) for $ a/\rho_{i} = 10 $ and $ B_{\textnormal{in}}/B_{\textnormal{v}} = \rho_{i}/a $. The dashed line shows the graph of the function $ B^{2}/8 \pi + \Delta \Pi_{rr} $.
  }\label{fig:(Bin,0_1)}
\end{figure*}



\begin{figure*}[tb]
  \centering
\includegraphics[width=0.47\linewidth]{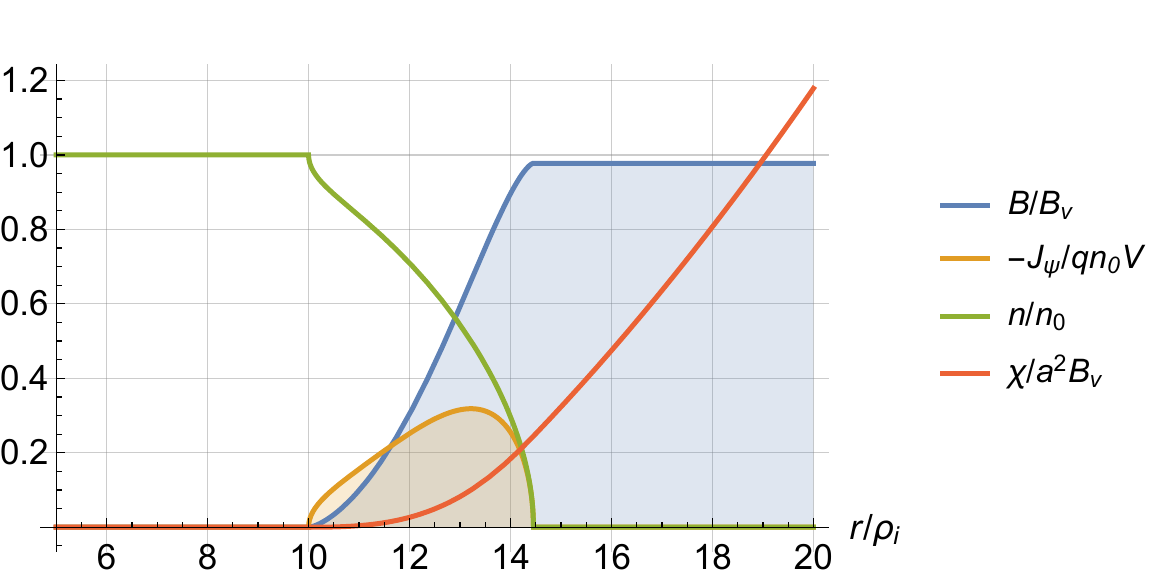}\hfil
\includegraphics[width=0.47\linewidth]{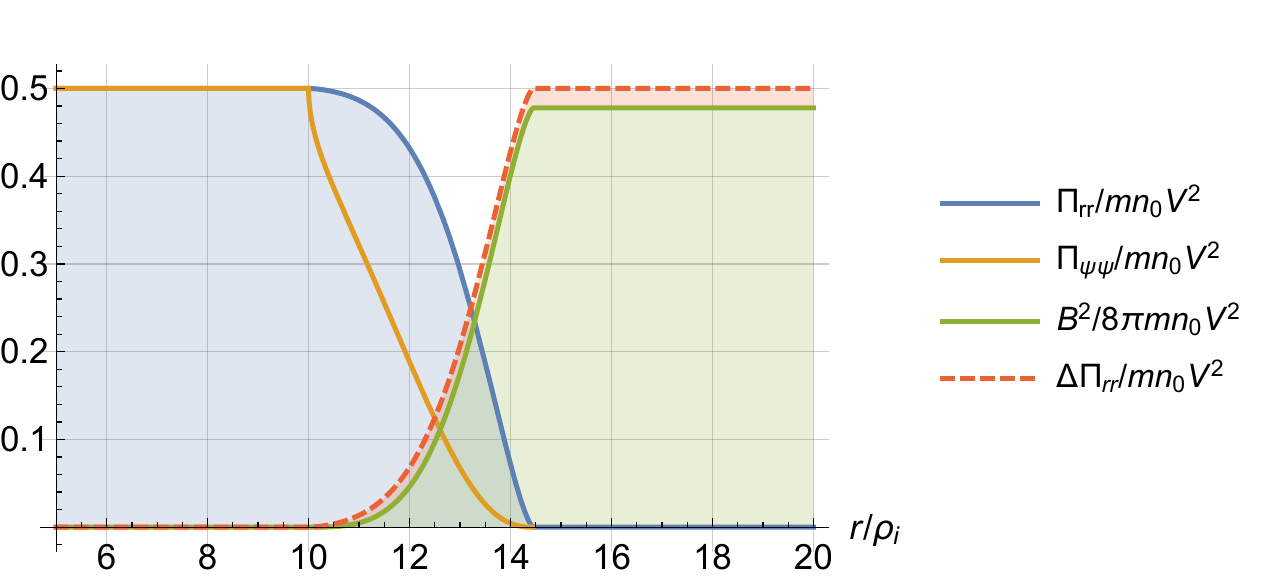}\hfil
\caption{
    Radial profiles of the magnetic field, current density, density (left) and components of the momentum flux tensor, magnetic field pressure (right) for $ a/\rho_{i} = 10 $ and $ B_{\textnormal{in}}/B_{\textnormal{v}} = 0 $. The dotted line shows the graph of the function $ B^{2}/8 \pi + \Delta \Pi_{rr} $.
  }\label{fig:(Bin,0)}
\end{figure*}

\begin{figure*}[tb]
  \centering
\includegraphics[width=0.47\linewidth]{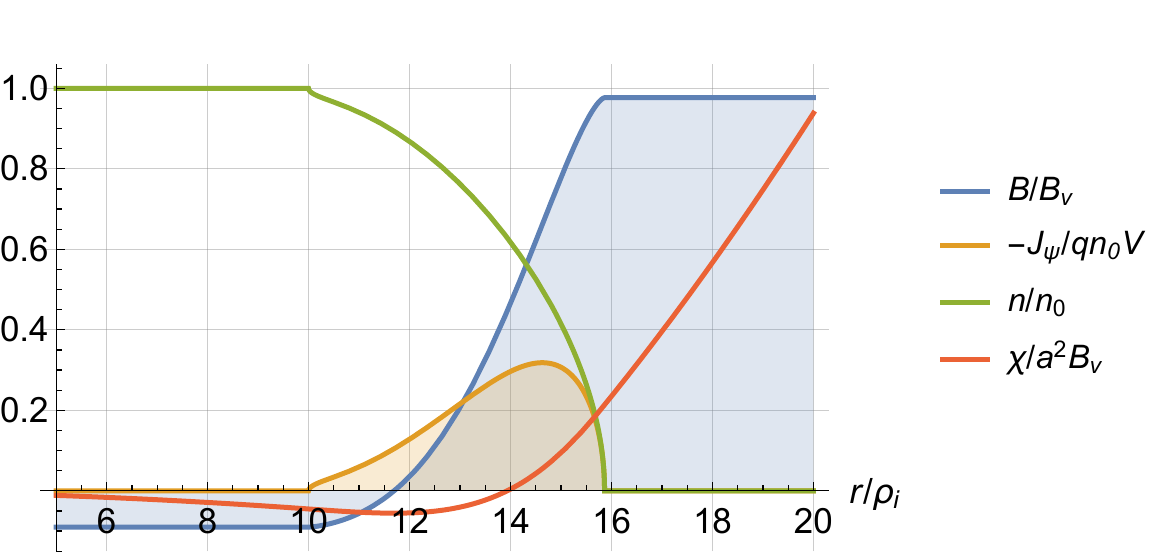}\hfil
\includegraphics[width=0.47\linewidth]{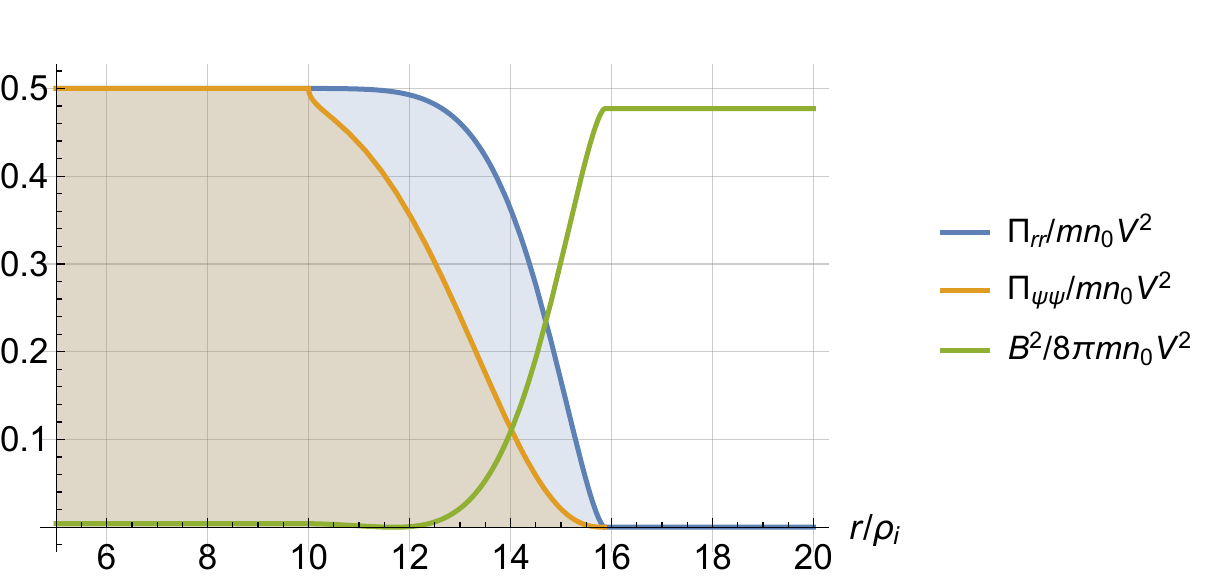}
\caption{
    Radial profiles of the magnetic field, current density, density (left) and components of the momentum flux tensor, magnetic field pressure (right) for $ a/\rho_{i} = 10 $ and $ B_{\textnormal{in}}/B_{\textnormal{v}} = - 0.9 \rho_{i}/a $.  The inverted field solution disappears when $ B_{\textnormal{in}}/B_{\textnormal{v}}  = - \rho_{i}/ a $.
  }\label{fig:(Bin,-0_09)}
\end{figure*}

%
The ratio $ B_{\textnormal{in}}/B_{\textnormal{v}} $ of the magnetic field $ B_{\textnormal{in}} $ on the inner side of the boundary layer to the vacuum field $ B_{\textnormal{ v}} $ along with the relation $ a/\rho_{i} $ are key dimensionless parameters of the problem. In the class of solutions corresponding to the case $ B_ {\textnormal{in}}/B_{\textnormal{v}} \geq 0 $, the magnetic flux $ \chi (r) $ turns out to be a monotonically growing function of radius $ r $. Given our definition of the function $G$ on the inner side of the boundary layer, $ G(a) = 0 $, it should be concluded that $ G <0 $ in the internal cavity of the diamagnetic bubble and $ G>0 $ in the boundary layer and outside the bubble. It means that
    \begin{equation}
    \label{6:14}
    J_{\psi}
    =
    -
    \frac{1}{\pi r}\,
    \sqrt{{r^2}-{(a-G)^2}}
    \end{equation}
in some neighborhood of the point $ r = a $ if $ r> a $ (second line in Eq.~\eqref{6:07}), and
    \begin{equation}
    \label{6:15}
    J_{\psi}=0
    \end{equation}
for all $ r <a $ (the third line Eq.~\eqref{6:07}). Zero current density at $ r <a $ means that inside the bubble the magnetic field is uniform and equal to $ B_{\textnormal{in}} $, and
    \begin{equation*}
    \chi(r) = \frac{1}{2} B_{\textnormal{in}}\,r^{2}
    .
    \end{equation*}
%
Together with the magnetic field, the ion density and pressure tensor components have a uniform profile, as shown in the figures \ref{fig:(Bin,0_5)},
\ref{fig:(Bin,0_1)},
\ref{fig:(Bin,0)} and \ref{fig:(Bin,-0_09)} for different values of $ B_{\textnormal{in}}/ B_{\textnormal{v}} $ from $ 0.5 $ to $ -0.009 $ and a fixed value of $ a/\rho_{i} = 10 $. A solution with a negative value of $ B_{\textnormal{in}} $ in the topology of the diamagnetic bubble disappears when $ B_{\textnormal{in}}/B_{\textnormal{v}} <- \rho_{i}/a $.

%
When analyzing the plots, it is noteworthy that the sum of the pressure of the ions and the magnetic field inside the bubble exceeds the pressure of the magnetic field outside, and this excess increases with decreasing bubble radius $ a $.
However, in the one-dimensional problem, the equality \eqref{4:07} holds, which guarantees the equality of the sum $ P_{xx} + B^{2}/8 \pi $ inside and the pressure of the vacuum magnetic field $ B_ {\textnormal{v}}^{2}/8 \pi $ outside the diamagnetic bubble, therefore it is clear that this effect is associated with the cylindrical geometry of the problem and the pressure anisotropy inside the boundary layer. Indeed, plasma equilibrium in the general case means the equality
    \begin{equation}
    \label{6:16}
    \Div\tensor{\Pi}
    =
    \frac{4\pi}{c}\left[
        \vec{J}\times \vec{B}
    \right]
    ,
    \end{equation}
%
which is obtained when calculating the moments of kinetic equilibrium and, therefore, is true even if the magnetohydrodynamic approximation is not used (see, for example, \cite {Kotelnikov2013Binom(eng)}). In the case under consideration, when the components of the momentum flux tensor $ \tensor {\Pi} $ depend only on the radial coordinate, Eq.~\eqref{6:16} gives
    \begin{equation}
    \label{6:17}
    \Pi_{rr}'(r)
    +
    \frac{\Pi_{rr}(r)-\Pi_{\psi\psi} (r)}{r}
    =
    - \frac{BB'}{4\pi}
    .
    \end{equation}
%
In a traditional approach to description of anisotropic plasma where  $ \Pi_{rr} = \Pi_{\psi \psi} = p_{\bot}$ Eq.~\eqref{6:17} reduces to the equality $ p_{\bot} + B^{2}/8 \pi = \const $. However, inside the boundary layer all three diagonal components of the momentum flux tensor ($ \Pi_{rr}$, $\Pi_{\psi \psi} $, $\Pi_{zz}$) are different and $ \Pi_ {rr}> \Pi_{\psi \psi} $. It seems that this case has not been previously investigated in detail in the scientific literature. Integrating Eq.~\eqref{6:17} and taking into account the fact that $ \Pi_{rr} = \Pi_{\psi\psi} $ for $ r <a $, we obtain the equality
    \begin{equation}
    \label{6:18}
    \Pi_{rr}
    +
    \Delta \Pi_{rr}
    +
    \frac{B^{2}}{8\pi}
    =
    \frac{1}{2}mn_{0}V^{2} + \frac{B_{\textnormal{in}}^{2}}{8\pi}
    ,
    \end{equation}
where
    \begin{equation*}
    \label{6:19}
    \Delta \Pi_{rr}(r)
    =
    \int_{0}^{r}
    \frac{\dif{r}}{r}
    \left(
        {\Pi_{rr}(r)-\Pi_{\psi\psi} (r)}
    \right)
    .
    \end{equation*}
%
In Fig.~\ref{fig:(Bin,0_5)},
\ref{fig:(Bin,0_1)},
and \ref{fig:(Bin,0)} the $ \Delta \Pi_{rr} $ curve is drawn over the magnetic field pressure graph $ B^{2}/8 \pi $ and shown as the sum $ \Delta \Pi_{rr } + B^{2}/8 \pi $, and the parts corresponding to $ B^{2}/8 \pi $ and $ \Delta \Pi_{rr} $ are shaded in different colors. We have verified numerically that the equality \eqref{6:18} is fulfilled with great accuracy.

We obtain a solution in the region $ r> a $ at a small distance $ x = r-a $ from the inner boundary of the surface layer $ r = a $. Near this boundary, the equation for the magnetic potential takes the form
    \begin{equation}
    \label{6:21}
    \chi''(x)
    =
    \frac{\sqrt{2B_{\textnormal{v}}a}}{\pi\rho_{i}}\,
    \sqrt{
        x + \frac{\chi(x)}{B_{\textnormal{v}}\rho_{i}}
    }.
    \end{equation}
Having made the replacement $ Q (x) = \chi (x) + B_{\textnormal{v}} \rho_{i} x $, its solution can be written in quadratures as an implicit function
    \begin{equation}
    \label{6:22}
    x
    =
    \frac{Q}{Q'(0)}\,
    {_2}F_1\left(
        \frac{1}{2},
        \frac{2}{3};
        \frac{5}{3};
        -\frac{
            4 \sqrt{2B_{\textnormal{v}}a}\, Q^{3/2}
        }{
            3 \pi  Q'(0)^2 \rho_i^{3/2}
        }
    \right)
    ,
    \end{equation}
%
where $ {_2} F_1 $ denotes hypergeometric function, and $ Q '(0) = B_{\textnormal{v}} \rho_{i} + B_{\textnormal{in}} a $ is the derivative of the function $ Q (x) $ at the point $ x = 0 $. Hence, in the limit $ x \to0 $ (which approximately corresponds to $ x \lesssim \rho_{i} $) we obtain
    \begin{equation}
    \label{6:23}
    B(x)=\frac{1}{a}\chi'(x)
    =
    B_{\textnormal{in}}
    +
    \frac{
        2 \sqrt{2B_v}
        \sqrt{B_v \rho_i+ B_{\textnormal{in}}a}
    }{
        3 \pi  \sqrt{a}
    }
    \left(
        \frac{x}{\rho_i}
    \right)^{3/2}
    .
    \end{equation}
%
In another limit, $ Q \to \infty $ (which roughly corresponds to $ x \gtrsim \rho_{i} $) we obtain
    \begin{equation}
    \label{6:24}
    B(x)=\frac{1}{a}\chi'(x)
    =
    \frac{ B_v }{ 18 \pi ^2 }
    \left(
        \frac{x-x_0}{\rho_i}
    \right)^3
    -
    \frac{\rho_i}{a}
    B_v
    ,
    \end{equation}
where
    \begin{equation}
    \label{6:25}
    x_{0}
    =
    -\frac{
        3^{5/3} \sqrt[6]{\pi } \Gamma
        \left({5}/{6}\right) \Gamma
        \left({5}/{3}\right) \rho_i
        \sqrt[3]{Q'(0)}
    }{
        2^{2/3} \sqrt[3]{B_{\textnormal{v}}a}
    }
    .
    \end{equation}
The first term in Eq.~\eqref{6:24} coincides with Eq.~\eqref{4:03} for the magnetic field in the flat boundary layer.



\section{Adiabatic invariant}\label{s7}


The first example we know of constructing an adiabatic invariant for charged particles, the trajectory of which in a magnetic field is not even remotely similar to the Larmor spiral, was undertaken by the author in  \cite{KotelnikovSchetnikov1987BINP_87_10(eng)}. The referred paper analyzed the motion of fast ions in an axially symmetric gas-dynamic trap with a relative plasma pressure of the order of unity, $ \beta \sim 1 $, and it was assumed that the transverse diameter of the orbit is not small compared with the plasma radius. Below in this section we follow the ideas of this publication.

%

To construct an adiabatic invariant corresponding to a slow variation in the magnetic field along the trap axis $ z $, one should find an unperturbed system in which the ion motion is periodic.

We recall first how the expression for the conventional adiabatic invariant is obtained, which is not quite justifiably identified with the magnetic moment $ \mu = mV_{\bot}^{2}/2B $.\footnote{
    As shown in \cite [\S12.5] {Jackson2001} and \cite [\S5.1] {Kotelnikov2013Binom(eng)}, the first adiabatic invariant is the magnetic field flux through the Larmor section of the trajectory of a charged particle in a magnetic field. It coincides with the magnetic moment only in the nonrelativistic limit.
}
In this case, the unperturbed system is a uniform constant magnetic field in which the trajectory of the charged particle is closed in a reference frame in which the longitudinal velocity of the particle is zero. Let us compute the integral
    \begin{equation}
    \label{7:01}
    I
    =
    \frac{1}{2\pi} \oint
    \vec{p}_{t}{\cdot }\dif{\vec{r}}
    \end{equation}
taken over a complete period of the motion, i.e. over the circumference of a circle in the present case with $\vec{p}_{t}$  being the projection of the generalized momentum on the plane perpendicular to $\vec{B}$. Substituting $\vec{p}_{t} = m\vec{v}_{t} + (e/c)\, \vec{A}$, we have:
    \begin{equation}
    \label{7:02}
    I = \frac{1}{2\pi} \oint \vec{p}_{t}{\cdot }\dif{\vec{r}}
    =
    \frac{1}{2\pi} \oint m\vec{v}_{t}{\cdot }\dif{\vec{r}}
    +
    \frac{e}{2\pi c} \oint \vec{A}{\cdot }\dif{\vec{r}}
    .
    \end{equation}
In the first term we note that $\vec{v}_{t}$ is constant in magnitude and directed along $\dif{\vec{r}}=\vec{v}_{t}\dif{t}$; we apply Stokes' theorem to the second term, write $\curl\vec{A} = \vec{B}$ and obtain\footnote{
    By inspecting the direction of motion of a charge along the orbit for a given direction of $\vec{B}$, we observe that it is counterclockwise if we look along $\vec{B}$. Hence the negative sign in the second term.
}
    \begin{equation}
    \label{7:03}
    I = r m v_{t} - \frac{e}{2c}Br^{2}
    =
    \frac{cm^{2}v_{t}^{2}}{2eB}
    =
    \frac{mv_{t}^{2}}{2\Omega}
    =
    \frac{e}{2c}Br^{2}
    ,
    \end{equation}
where $\Omega =eB/mc$. According to general theory of adiabatic invariants \cite[\S49]{LLI(eng)}, the magnitude of $I$ approximately remains constant when the magnetic field slowly varies in time or in space. From this we see that, for slow variation of $B$, the tangential momentum $mv_{t}$ varies proportionally $\sqrt{B}$. 
Note that the formula $ I = {mv_{t}^{2}}/{2\Omega} $ is true only in the nonrelativistic limit, but the alternative form $ I = \left ({e}/{2c} \right) Br^{2} $ is always valid.


In general the integrals $\oint p \dif{q}$, taken over a period of the particular coordinate $q$, are adiabatic invariants. In the present case the periods for the two coordinates in the plane perpendicular to $\vec{B}$ coincide, and the integral $I$ which we have written is the sum of the two corresponding adiabatic invariants, $\oint p_{x} \dif{x}$ and $\oint p_{y} \dif{y}$. However, each of these invariants individually has no special significance, since it depends on the (non-unique) choice of the vector potential of the field.


Let us return to the search for the adiabatic invariant for a diamagnetic bubble. From what has been said above, it should be clear that an unperturbed system is such a system where the motion is strictly periodic, that is, the ion trajectory is closed. If we neglect the dependence of the bubble parameters on the coordinate $ z $ along its axis, then it is obvious that the ion trajectory will be closed in a coordinate system that rotates around the $ z $ axis with some specially selected frequency $ \omega $, as shown in Fig.~\ref{fig:MotionInCylinder}, and moves along the $ z $ axis with the longitudinal velocity of the particle. The integral in Eq.~\eqref{7:01} must be calculated in the coordinate system, where the trajectory is closed. Thus, we need to find an expression for the generalized angular momentum in a rotating coordinate system.

%
It is known that in the fixed coordinate system\footnote{For brevity, in this section we do not write the third coordinate $ z $ along the direction of the magnetic field.} $ \vec {r} = (x, y) $ canonical momentum $ \vec {p } = (p_{x}, p_{y}) $ is associated with the velocity vector $ \vec {v} = (v_{x}, v_{y}) = (\dot x, \dot y) $ and the vector potential $ \vec {A} = (A_{x}, A_{y}) $ by the formula
    \begin{equation}
    \label{7:04}
    \vec{p} = m\vec{v} + \frac{e}{c}\,\vec{A}
    ,
    \end{equation}
and the Hamiltonian has the form
    \begin{equation}
    \label{7:05}
    \Hamiltonian = \frac{1}{2m}\left(
        \vec{p}-\frac{e}{c}\,\vec{A}
    \right)^{2}
    .
    \end{equation}

To convert coordinates and momenta into a Cartesian coordinate system $ (X, Y) $, which rotates at a frequency of $ \omega $, we use the generating function
    \begin{multline}
    \label{7:06}
    \Phi
    =
    P_X \left(
        x \cos (\omega t)+y \sin (\omega t )
    \right)
    \\
    +
    P_Y \left(
        y \cos (\omega t)-x \sin (\omega t)
    \right)
    ,
    \end{multline}
that depends on the old coordinates $ (x, y) $ and the new momenta $ (P_{X} , P_{Y}) $. Transformation to new coordinates is performed according to the formulas
    \begin{gather}
    \label{7:07}
    \begin{gathered}
    X
    = \parder{\Phi}{P_{X}}
    = \cos(\omega t)\,x  + \sin(\omega t)\,y,
    \\
    Y
    = \parder{\Phi}{P_{Y}}
    = -\sin(\omega t)\,x + \cos(\omega t)\,y
    .
    \end{gathered}
    \end{gather}
Differentiating them with respect to time, we find the velocity transformation rule:
    \begin{gather}
    \label{7:08}
    \begin{gathered}
    V_{X}
    =
    \cos(\omega t)\,v_{x} + \sin(\omega t)\,v_{y}
    + \omega Y
    ,
    \\
    V_{Y}
    =
    -\sin(\omega t)\,v_{x} + \cos(\omega t)\,v_{y}
    -\omega X
    .
    \end{gathered}
    \end{gather}
Old impulses $ (p_{x}, p_{y}) $ are expressed in terms of new impulses $ (P_{X}, P_{Y}) $ by the equations
    \begin{gather}
    \label{7:09}
    \begin{gathered}
    p_x
    = \parder{\Phi}{x}
    = P_X \cos (\omega t)-P_Y \sin (\omega t)
    ,\\
    p_y
    = \parder{\Phi}{y}
    = P_X \sin (\omega t)+P_Y \cos (\omega t)
    .
    \end{gathered}
    \end{gather}
New Hamiltonian
    \begin{equation}
    \label{7:11}
    \Hamiltonian' = \Hamiltonian + \parder{\Phi}{t}
    \end{equation}
must be expressed in terms of new coordinates and momenta. By transforming the components of the vector potential,
    \begin{gather}
    \label{7:12}
    \begin{gathered}
    A_{X}
    = \cos(\omega t)\,A_{x}  + \sin(\omega t)\,A_{y},
    \\
    A_{Y}
    = -\sin(\omega t)\,A_{x} + \cos(\omega t)\,A_{y}
    ,
    \end{gathered}
    \end{gather}
we obtain
    \begin{multline}
    \label{7:14}
    \Hamiltonian'
    =
    \frac{1}{2m}\left(
        P_{X} - \frac{e}{c}A_{X}
    \right)^{2}
    +
    \frac{1}{2m}\left(
        P_{Y} - \frac{e}{c}A_{Y}
    \right)^{2}
    \\
    -
    \omega \left(
        X P_{Y} - Y P_{X}
    \right)
    .
    \end{multline}
The velocity and the generalized momentum in the rotating coordinate system are related by the Hamilton equations
    \begin{gather}
    \label{7:15}
    \begin{gathered}
    V_{X} = \dot{X} = \parder{\Hamiltonian'}{P_{X}}
    ,
    \qquad
    V_{Y} = \dot{Y} = \parder{\Hamiltonian'}{P_{X}}
    .
    \end{gathered}
    \end{gather}
It follows from them that
    \begin{gather}
    \label{7:16}
    \begin{gathered}
    P_{X} = m V_{X} + \frac{e}{c}\,A_{X} - m \omega Y
    ,\\
    P_{Y} = m V_{Y} + \frac{e}{c}\,A_{Y} + m \omega X
    .
    \end{gathered}
    \end{gather}

In order to calculate the rotation frequency $ \omega $ of the coordinate system in which the ion trajectory is closed, it is convenient to convert the Cartesian coordinates $ (X, Y) $ to the polar coordinates $ (R, \Psi) $. Such a conversion is carried out using the generating function
    \begin{equation}
    \label{7:21}
    \Phi'
    =
    P_R\, \sqrt{X^2+Y^2}
    +
    P_{\Psi } \arctan\left(Y/X\right)
    \end{equation}
and equations
    \begin{gather}
    \label{7:22}
    \begin{gathered}
    R = \parder{\Phi'}{P_{R}}
    ,\qquad
    \Psi = \parder{\Phi'}{P_{\Psi }}
    ,\\
    P_{X} = \parder{\Phi'}{X}
    ,\qquad
    P_{Y} = \parder{\Phi'}{Y}
    .
    \end{gathered}
    \end{gather}
Solving them, we find
    \begin{gather}
    \label{7:23}
    \begin{gathered}
    X = R \cos(\Psi )
    ,\qquad
    P_{R} = P_{X}\cos(\Psi ) + P_{Y}\sin(\Psi )
    ,
    \\
    Y = R \sin(\Psi )
    ,\qquad
    P_{\Psi } =
        - R P_{X}\sin(\Psi ) + R P_{Y}\cos(\Psi )
    .
    \end{gathered}
    \end{gather}
Since $ \Phi'$ does not depend explicitly on time, to find the Hamiltonian in polar coordinates, it is sufficient to express $ X $, $ Y $, $ P_{X} $, $ P_{Y}$ in Eq.~\eqref{7:14} through $ R $, $ \Psi $, $ P_{R} $, $ P_{\Psi} $. Transforming also the components of vector potential,
    \begin{gather}
    \label{7:24}
    \begin{gathered}
    A_{R} = A_{X}\cos(\Psi ) + A_{Y}\sin(\Psi )
    ,\\
    A_{\Psi } =
        - A_{X}\sin(\Psi ) + A_{Y}\cos(\Psi )
    ,
    \end{gathered}
    \end{gather}
we write the Hamiltonian in polar coordinates:
    \begin{equation}
    \label{7:25}
    \Hamiltonian'
    =
    \frac{1}{2m}\left(
        P_{R} - \frac{e}{c}\,A_{R}
    \right)^{2}
    +
    \frac{1}{2m R^{2}}\left(
        P_{\Psi } - \frac{e}{c}\,RA_{\Psi }
    \right)^{2\emph{}}
    -
    \omega P_{\Psi }
    .
    \end{equation}
In a similar way, one could transform the Cartesian coordinates $ (x, y) $ to the polar coordinates $ (r, \psi) $ in a fixed system. It is easy to verify that
    \begin{gather}
    \label{7:26}
    \begin{gathered}
    r=R,\qquad \psi = \Psi + \omega t
    ,\\
    p_{r} = P_{R}, \qquad p_{\psi} = P_{\Psi }
    .
    \end{gathered}
    \end{gather}
From the Hamiltonian equations
    \begin{gather}
    \label{7:27}
    \begin{gathered}
    \dot{R} = \parder{\Hamiltonian'}{P_{R}}
    ,\qquad
    \dot{\Psi } = \parder{\Hamiltonian'}{P_{\Psi }}
    ,\\
    \dot{P}_{R} = - \parder{\Hamiltonian'}{R}
    ,\qquad
    \dot{P}_{\Psi} = - \parder{\Hamiltonian'}{\Psi}
    ,
    \end{gathered}
    \end{gather}
we obtain
    \begin{gather}
    \label{7:28}
    \begin{gathered}
    P_{R} = m \dot{R} + \frac{e}{c}\,A_{R}
    ,\qquad
    P_{\Psi } = m R^{2}\dot{\Psi }
    + \frac{e}{c}\, R\, A_{\Psi } + m R^{2} \omega
    ,
    \end{gathered}
    \end{gather}
where $P_{\Psi}=\const$ since $\tparder{\Hamiltonian'}{\Psi}=0$.

%
Below we move on to the notation $ r $, $ p_{r} $, $ p_{\psi} $ instead of $ R $, $ P_{R} $, $ P_{\Psi} $, which are equivalent due to \eqref{7:26}. We also denote by $ \varepsilon = \Hamiltonian = \Hamiltonian '+ \omega P_ {\Psi} = \const $ the ion energy in a fixed coordinate system, and  take into account that, due to the axial symmetry of the problem
$ A_{R} = A_{r} = 0 $, $ A_{\Psi} = A_{\psi} = \chi/r $. Combining Eqs.~\eqref{7:25} and~\eqref{7:28} we express the radial velocity $ \dot {r} $ as a function of $ r $ and the motion constants:
    \begin{equation}
    \label{7:31}
    \dot{r}
    =
    \frac{p_{r}}{m}
    =
    \frac{1}{m}\,
    \sqrt{
        2m\varepsilon
        -
        \left(
            p_{\psi} - (e/c)\,\chi
        \right)^{2}/r^{2}
    }
    \,
    .
    \end{equation}
Integrating this equation between the trajectory points closest to the axis and farthest from the axis, $ r_{\min} $ and $ r_{\max} $, where $ \dot {r} = 0 $, we can calculate the period of movement along the trajectory:
    \begin{equation}
    \label{7:32}
    T = 2\int_{r_{\min}}^{r_{\max}}
    \frac{\dif{r}}{\dot{r}}
    .
    \end{equation}
Integrating another equation
    \begin{equation}
    \label{7:33}
    \dot{\Psi }
    =
    \left(
        p_{\psi} - (e/c)\,\chi
    \right)/mr^{2}
    - \omega
    ,
    \end{equation}
we take into account that the change in the angle $ \Psi $ during the period of motion is zero,
    \begin{equation}
    \label{7:34}
    \Delta\Psi
    =
    2 \int_{r_{\min}}^{r_{\max}}
    \dot{\Psi }\,\frac{\dif{r}}{\dot{r}}
    =
    0
    ,
    \end{equation}
since the trajectory is closed in the rotating system.
From here we get the formal expression for the frequency of rotation:
    \begin{equation}
    \label{7:35}
    \omega
    =
    \frac{2}{T}
    \int_{r_{\min}}^{r_{\max}}
    \frac{
        \left(
            p_{\psi} - (e/c)\,\chi
        \right)/r^{2}
    }{
        \sqrt{
            2m\varepsilon
            -
            \left(
                p_{\psi} - (e/c)\,\chi
            \right)^{2}/r^{2}
        }
    }
    \dif{r}
    .
    \end{equation}
Rewriting Eq.~\eqref{7:01} as
    \begin{equation}
    \label{7:36}
    I
    =
    \frac{1}{2\pi}\oint
    \left[
        P_{R}\dif{R} + P_{\Psi}\dif{\Psi}
    \right]
    ,
    \end{equation}
we note that $ \oint P_{\Psi} \dif {\Psi} = P_{\Psi} \oint \dif {\Psi} = 0 $, since the constant value $ P_{\Psi} $ can be removed from under sign of the integral, and $ \oint \dif {\Psi} = 0 $, since the integral of the total differential along a closed path is equal to zero. Therefore
    \begin{equation}
    \label{7:37}
    I
    =
    \frac{1}{2\pi}\oint P_{R}\dif{R}
    =
    \frac{1}{\pi}\int_{r_{\min}}^{r_{\max}} p_{r}\dif{r}
    ,
    \end{equation}
and $ p_{r} $, according to \eqref{7:31}, does not depend on $ \omega $, as it should be for a true invariant.

%
%
%

%
As an example, we consider the limiting case when the magnetic field is completely absent inside the diamagnetic bubble, but rapidly grows in the boundary layer at $ r> a $. Then we can assume that the ion trajectory lies entirely in the region $ r \leq a $, where $ \chi = 0 $. Substituting $\varepsilon = mV_{\bot}^{2}/2 $ and $ p_{\psi} = mV_{\bot} a \sin (\alpha) $ in the equation \eqref{7:31}, we find that $ r_{\min} = a \left | \sin (\alpha) \right | $, $ r_{\max} = a $. Movement period
    \begin{equation}
    \label{7:41}
    T = \frac{2a\cos(\alpha )}{V_{\bot}}
    ,
    \end{equation}
angle of rotation
    \begin{equation}
    \label{7:42}
    \Delta \psi
    = \left(\pi - 2\left|\alpha\right| \right)\sign(\alpha)
    \end{equation}
and frequency of rotation
    \begin{equation}
    \label{7:43}
    \omega
    =
    \frac{\Delta\psi}{T}
    =
    \sign(\alpha)\,
    \frac{\pi-2\left|\alpha\right|}{2\cos(\alpha )}\,\frac{V_{\bot}}{a}
    \end{equation}
%
can be calculated either using Eqs.~\eqref{7:32} and~\eqref{7:35}, or using simple geometric treatment, as in Fig.~\ref{fig:CoordinatesInCylinder}. The angle $ \alpha $ in the above formulas varies in the interval $ - \pi/2 <\alpha <+ \pi/2 $.

%
Finally, by integrating in Eq.~\eqref{7:37} from $ r_{\min} = a \left| \sin (\alpha) \right | $ to $ r_{\max} = a $ with taking into account Eq.~\eqref{7:31} for $ \chi = 0 $, we calculate the part of the adiabatic invariant associated with the motion inside the diamagnetic bubble:
    \begin{equation}
    \label{7:44}
    I_{0}
    =
    \frac{m V_{\bot} a}{2\pi} \left[
        2 \cos (\alpha )
        -
        (\pi -2 \left|\alpha\right| )
        \left|\sin (\alpha )\right|
    \right]
    .
    \end{equation}
%
The correction associated with the motion in the boundary layer is calculated in the limit $ \rho_{i} \to0 $, assuming that the particle penetrates into this layer at a very small distance $ x \ll a $, where the magnetic potential is approximately described by the formula
    \begin{gather}
    \label{7:45}
    (e/c)\left[
        \chi(a+x)-\chi(a)
    \right] \approx
    a G
    =
    \frac{2mVa}{9\pi^{2}}\left(
        \frac{x}{\rho_{i}}
    \right)^{4}
    ,
    \end{gather}
that follows from Eq.~\eqref{4:03}. Accordingly, the desired correction is expressed by the integral
    \begin{equation}
    \label{7:46}
    I_{1}
    =
    \frac{mV\rho_{i}}{2\pi} \int_{0}^{x_{\max}}
    \sqrt{
        1 - \left(
            \sin(\alpha )
            -
            \frac{2}{9\pi^{2}}
            \left( \frac{x}{\rho_{i}} \right)^{4}
        \right)^{2}
    }
    \frac{\dif{x}}{\rho_{i}}
    ,
    \end{equation}
where $ x_{\max} $ is the root of the equation
    \begin{equation}
    \label{7:47}
    1+\sin(\alpha )
    =
    \frac{2}{9\pi^{2}}
    \left( \frac{x}{\rho_{i}} \right)^{4}
    .
    \end{equation}
The result of integration is expressed through the hypergeometric function:
    \begin{multline}
    \label{7:48}
    I_{1}
    =
    \frac{
        3 \sqrt{3}  \Gamma\left(\frac{5}{4}\right)
        m V_{\bot}\rho_{i}
    }{
        5 \sqrt[4]{2} \Gamma \left(\frac{7}{4}\right)
        \sqrt[4]{1+\sin(\alpha)}\,
        \sqrt{1-\sin(\alpha )}
    }
    \times
    \\
    \times
    \left[
        2 \sin (\alpha ) \,
        {_2F_1}\left(
            \frac{1}{2},\frac{5}{4};\frac{3}{4};
            -\frac{1+\sin (\alpha )}{1-\sin (\alpha )}
        \right)
    \right.
    \\
    \left.
        +
        \left(1-\sin(\alpha )\right)
        {_2F_1}\left(
            \frac{1}{4},\frac{1}{2};\frac{3}{4};
            -\frac{1+\sin (\alpha )}{1-\sin (\alpha )}
        \right)
    \right]
    .
    \end{multline}
\begin{figure}
  \centering
  \includegraphics[width=\linewidth]{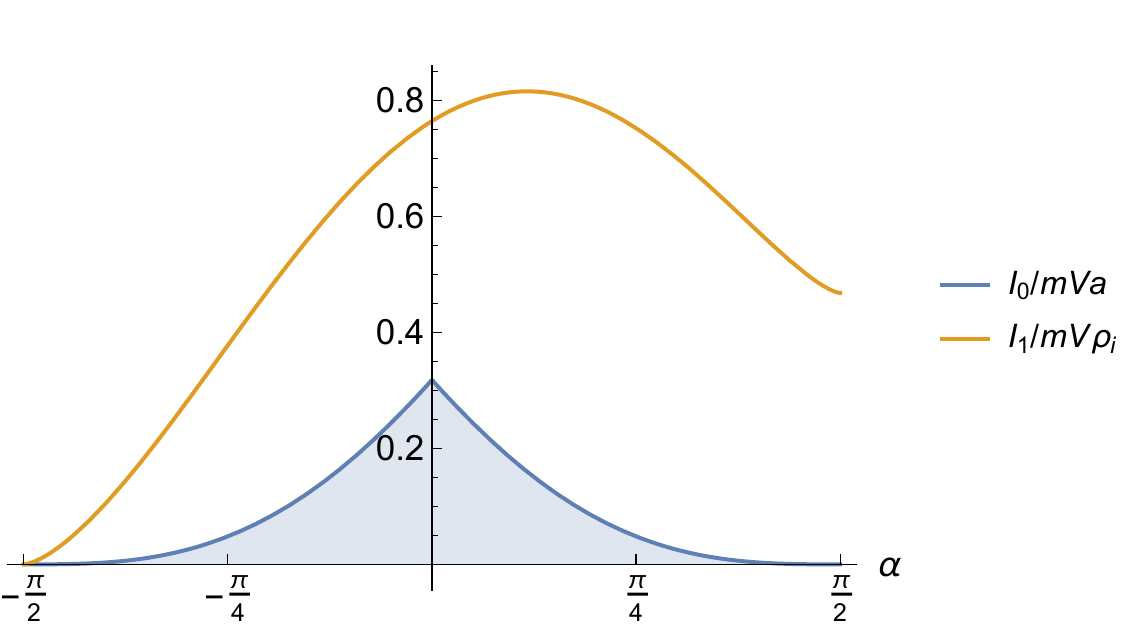}
  \caption{
    Adiabatic invariant in the approximation $ \rho_{i} \ll a $.
  }\label{fig:AdiabaticI}
\end{figure}
The graphs of the functions \eqref{7:44} and \eqref{7:48} are drawn in Fig.~\ref{fig:AdiabaticI}.

%
The adiabatic invariant $ I = I_{0} + I_{1} $ binds three parameters: $ a $, $ \alpha $ and $ V_{\bot} $ unlike the magnetic moment $ \mu = mV_ { \bot}^{2}/2B $, which binds only two parameters: $ V_{\bot} $ and $ B $. To determine how the transverse and longitudinal speeds change, $ V_{\bot} $ and $ V_{ \|} $, in an adiabatic trap two conservation laws are sufficient: $ \varepsilon = m \left (V_{\bot}^ {2} + V_{\|}^{2} \right)/2 = \const $ and $ \mu = mV_{\bot}^{2}/2B = \const $. Three conservation laws are needed in the diamagnetic trap, and the third law is the conservation of the generalized azimuthal moment $ p_{\psi} = mV_{\bot} a \sin (\alpha) = \const $. In the limit $ \rho_{i} \ll a $ the ratio $ p_{\psi} $ and the adiabatic invariant $ I \approx I_{0} $ depends only on the angle $ \alpha $, therefore it is preserved during adiabatic motion. Consequently, the product $ V_{\bot} a = \const $ is also preserved. The adiabatic invariance of the angle $ \alpha $ and the product $ V_{\bot} a $ is proved in Appendix \ref{A1} by direct integrating the equations of motion.


Note that the formula \eqref{7:44} (in other notation) was previously obtained by Ivan Chernoshtanov \cite{Chernoshtanov2020Arxiv_2002_03535}, but he did not notice that the angle of incidence $\alpha $ and the product $aV$ also possess the properties of the adiabatic invariant. The second remark concerns the accuracy of Eq.~\eqref{7:44}. As can be seen by comparing the contributions to the adiabatic invariant of the interior of the diamagnetic bubble and the boundary layer, i.e. $ I_ {0} $ and $ I_ {1} $, the accuracy is very limited. For example, with $ a /\rho_ {i} = 10 $, the correction $ I_ {1} $ reaches 30\% of $ I_ {0} $.

The constancy of $ V_{\bot} a $ along with the conservation of energy $ \varepsilon $ allows us to find how $ V_{\bot} $ and $ V_{\|} $ change depending on the radius of the diamagnetic bubble $ a $. In particular, if in a section with a maximum radius of $ a_{\max} = a_{0} $ the particle had a velocity with components $ V_{\bot0} $, $ V_{\| 0} $, a part of a diamagnetic trap with radius
    \begin{equation}
    a
    >
    a_{0}\,\sqrt{\frac{V_{\bot0}^{2}}{V_{\bot0}^{2}+V_{\|0}^{2}}}
    .
    \end{equation}
Thus, provided that the adiabatic invariant is preserved, the particle is held in a diamagnetic trap if
    \begin{equation}
    \frac{V_{\bot0}^{2}}{V_{\bot0}^{2}+V_{\|0}^{2}}
    >
    \frac{a_{\min}^{2}}{a_{\max}^{2}}
    ,
    \end{equation}
where $ a_{\min} $ is the radius of the neck of the trap.

For comparison, recall the condition of adiabatic confinement in the Budker-Post mirror trap. Within the adiabatic approximation, a charged particle is confined in such a mirror cell if
    \begin{equation}
    \frac{V_{\bot0}^{2}}{V_{\bot0}^{2}+V_{\|0}^{2}}
    >
    \frac{B_{\min}}{B_{\max}}
    ,
    \end{equation}
%
where $ B_{\min} $ and $ B_{\max} $ are the minimum and maximum values of the magnetic field on the magnetic line along which the particle moves, and the component velocities $ V_{\bot0} $ and $ V_{\| 0} $ refer to the cross section of the mirror cell in which $ B = B_{\min} $. This criterion essentially requires that the magnetic moment of the particle $ \mu = mV _ {\bot}^{2}/2B $ be sufficiently large. However, the adiabatic invariant is not an exact integral of the motion. As shown by numerical simulation performed by Ivan Chernoshtanov \cite{Chernoshtanov2020Arxiv_2002_03535}, in the diamagnetic trap the adiabatic invariant $ I $ easily breaks because of the presence of nonparaxial region where the radius of the inner cavity with the zero magnetic field is quickly reduced (see~\cite{KhristoBeklemishev2019PlasFusRes_14_2403007}), as well as because of small but unavoidable ripple of magnetic field in the homogeneous part of the diamagnetic bubble. A criteria of adiabaticity of particle motion in the diamagnetic trap with smooth and corrugated magnetic field are discussed in \cite{Chernoshtanov2020Arxiv_2002_03535}.

\section{Absolute confinement}\label{s8}

Absolute is called such a confinement of a charged particle by an external electromagnetic field in a limited isolated region of space, the way out of which is impossible without violating the \emph {exact} integrals of motion (see \cite{,MorozovSolovyev1963VTP2(eng)}). The exact integrals of motion in a trap with an axially symmetric magnetic field are the energy
    \begin{equation}
    \label{8:01}
        \varepsilon=\tfrac{1}{2}m{v}^2
    \end{equation}
and generalized azimuthal moment (compare \eqref{6:03a})
    \begin{equation}
    \label{8:02}
    p_{\psi}
    =
    mrv_{\psi}
    + \frac{e}{c}\,\chi(r,z)
    ,
    \end{equation}
and it is assumed that in the cylindrical coordinate system the vector potential is calibrated so that it has only the azimuthal component $ A_\psi (r, z) = \chi (r, x)/r $. The Hamiltonian of a charged particle in such a field is
    \begin{equation}
    \Hamiltonian =\frac{p_r^2}{2m} +\frac{p_{z}^2}{2m}
    +
    \frac{1}{2mr^2}\left(
        p_{\psi}-\frac{e}{c}\,\chi
    \right)^2
    ,
    \end{equation}
where $ p_{r} = mv_r $ and $ p_{z} = mv_{z} $ denote the radial and axial components of the particle momentum. It describes the motion in two-dimensional potential
    \begin{equation}
    U(r,z)=\frac{1}{2mr^2}\left(
        p_{\psi} - \frac{e}{c}\,\chi(r,z)
    \right)^2
    .
    \end{equation}
The contours $ U = \const $ on the $ rz $ plane limit the region of space accessible to a particle with a given energy. If the contours are not closed, the particle can escape from the trap. Absolute confinement corresponds to closed contours around local minima $ U = U_{\min} $, where the first derivatives vanish,
    \begin{equation}
    \label{8:05}
        \parder{U}{r}=0
        ,
        \qquad
        \parder{U}{z}=0
        ,
    \end{equation}
and the second derivatives satisfy the conditions
    \begin{equation}
    \label{8:06}
        \parder{^2U}{r^2}>0
        ,
        \qquad
        \parder{^2U}{z^2}>0
        ,
        \qquad
        \parder{^2U}{r^2}\,\parder{^2U}{z^2}> \left(\parder{^2U}{r\partial z}\right)^2
        .
    \end{equation}
They are separated from open contours by a separatrix that passes through the saddle point, where the first derivatives also vanish, but one of the conditions \eqref{8:06} is not satisfied.

First, we recall the result of finding the region of absolute confinement in the paraxial mirror trap \cite [\S1.5] {MorozovSolovyev1963VTP2(eng)}. In such a trap, approximately
    \begin{equation}
    \label{8:07}
    \chi(r,z)\approx \tfrac{1}{2}B_0(z)\,r^{2}
    ,
    \end{equation}
where $ B_0 (z) $ is the magnetic field on the axis of the trap. As a result of simple calculations, we find that for $ ep_\psi> 0 $ the equations \eqref{8:05} have the solution $ r = \sqrt {2cp_\psi/ eB_0 (z)} $ for any $ z $, however, it corresponds to a “gorge” $ U = 0 $, which extends to infinite values of $ z $.

For $ ep_\psi <0 $, the equations \eqref{8:05} have a solution $ r = \sqrt {-2cp_\psi/eB_0 (z)} $ for $ z $ such that $ \tparder {B_0}{ z} = 0 $. At the minimum $ B_0 = B_{\min} $ there is a local minimum of the effective potential
    \begin{equation}
    \label{8:08}
    U_{\min}=-ep_{\psi}B_{\min}/{mc}
    .
    \end{equation}
A separatrix passes throught the maximum $ B_0 = B_{\max} $, which corresponds to the value of the effective potential
    \begin{equation}
    \label{8:09}
    U_\textnormal{sep} = -ep_{\psi}B_{\max}/{mc}
    .
    \end{equation}
Particles with energy $ m \left (v_r^2 + v_{z}^2 \right)/2 + U <U_ \textnormal{sep} $ are trapped inside the region that includes the minimum of $ U $ and bounded by the separatrix. Since, in fact, $ U = mv_{\psi}^2/2 $, sufficient condition for holding a particle in a trap can be written as
    \begin{equation}
    \label{8:10}
    \varepsilon < -ep_{\psi}B_{\max}/mc
    ,
    \end{equation}
where the energy is $ \varepsilon = m \left (v_r^2 + v_\psi^2 + v_ {z}^2 \right)/2 $ and the generalized azimuthal momentum $ p_{\psi} = mrv_{\psi} + ( e/2c) \, r^2B_0 (z) $, being exact integrals of motion, are uniquely determined by the initial velocity of the particle and its initial coordinates.

%
Let the particle initially be in the plane of the minimum of magnetic field, where $ B_{0} (0) = B_{\min} $ at the point with coordinates (see Fig.~\ref{fig:CoordinatesInCylinder})
    \begin{equation*}
    x = R_{0}\cos\psi - \rho_{0}\sin\theta
    ,\quad
    y = R_{0}\sin\psi + \rho_{0}\cos\theta
    ,\quad
    z = 0
    \end{equation*}
and had velocity
    \begin{equation*}
    v_{x} = v_{\bot0} \cos\theta
    ,\qquad
    v_{y} = v_{\bot0} \sin\theta
    ,\qquad
    v_{z} = v_{\|0}
    ,
    \end{equation*}
and $ v_{\bot0} = \rho_{0} \Omega_{0} $, $ \Omega_{0} = eB_{\min}/mc $. Then
    \begin{gather*}
    p_{\psi} = m\left(
        v_{y}x - v_{x}y
    \right)
    +
    \frac{m}{2}\left(
        x^{2} + y^{2}
    \right)  \Omega_{0}
    =
    \frac{m}{2}\left(
        R_{0}^{2} - \rho_{0}^{2}
    \right)\Omega_{0}
    ,\\
    \varepsilon = \frac{m}{2}\left(
        v_{x}^{2} + v_{y}^{2} + v_{z}^{2}
    \right)
    =
    \frac{m}{2}\left(
        v_{\bot0}^{2} + v_{\|0}^{2}
    \right)
    \end{gather*}
and the condition \eqref{8:10} is reduced to
    \begin{equation}
    \label{8:15}
    \frac{v_{\bot0}^{2}}{B_{\min}}
    >
    \frac{v_{\bot0}^{2} + v_{\|0}^{2}}{B_{\max}}
    +
    \frac{R_{0}^{2}\Omega_{0} ^{2}}{B_{\min}}
    .
    \end{equation}
It can be seen from this that the absolute confinement region is located inside the adiabatic confinement region
    \begin{equation}
    \label{8:16}
    \frac{v_{\bot0}^{2}}{B_{\min}}
    >
    \frac{v_{\bot0}^{2} + v_{\|0}^{2}}{B_{\max}}
    ,
    \end{equation}
%
coinciding with it at $ R_{0} = 0 $ for particles whose leading center moves along the axis of the system. It is also clear that the Larmor radius of the particle $ \rho_{0} = v_{\bot0}/\Omega_{0} $ should exceed the distance $ R_{0} $ from the axis of the system to the leading center, that is, the path should bypass the axis.

%
Let us return to the discussion of the diamagnetic trap. Now it cannot be argued that only such particles whose trajectory bypasses the axis of the system are absolutely trapped. From Fig.~\ref{fig:MotionInCylinder} it can be seen that in a fixed frame of reference (the upper row in Fig.~\ref{fig:MotionInCylinder}), the axis of the system is bypassed by particles with any sign of $ p_{\psi} $, whereas in the accompanying reference frame (bottom row) there are no trajectories that span the axis of the system. On the other hand, the assertion that there is no absolute confinement region in the case of $ ep_{\psi}> 0 $ remains valid. In the case of $ ep_{\psi} <0 $, the coordinates of the singular points are determined from the equations
    \begin{gather}
    \label{8:17}
    \left(c/e\right)p_{\psi}-\chi(r,z) + \parder{\chi(r,z)}{r} = 0
    ,\quad
    \parder{\chi(r,z)}{z} = 0,
    \end{gather}
%
which are obtained from Eq.~\eqref{8:05}. The second of them means that at such a point the radial component of the magnetic field is equal to zero. For the $ \chi $ function near the median plane of the diamagnetic bubble, the formula \eqref{7:45} can be used, while the paraxial approximation \eqref{8:07} is applicable near the magnetic mirror. Accordingly, the formula \eqref{8:09} for $ U_{\textnormal{sep}} $ is still valid, and with it the condition of absolute confinement \eqref{8:10}. For the minimum value of potential energy, instead of \eqref{8:08} in the approximation \eqref{7:45}, the following expression is obtained:
    \begin{equation}
    \label{8:18}
    U_{\min}
    =
    \frac{p_{\psi}^{2}}{2ma^{2}}
    \left(
        1 + \frac{x_{\max}}{4a}
    \right)^{2}
    \approx
    \frac{p_{\psi}^{2}}{2ma^{2}}
    ,
    \end{equation}
where
    \begin{equation}
    x_{\max}
    =
    \sqrt[3]
    {
        \frac{ 9\pi^{2}cp_{\psi} }{ 2a^{2}eB_{v} }
    }
    \,
    \rho_{i}
    \end{equation}
%
is the positive root of the first of the equations \eqref{8:17}. The refinement of the expression for $ U_{\min} $ does not affect the condition of absolute confinement in the form \eqref{8:10}, as follows from the reasoning before Eq.~\eqref{8:10}. However, another form \eqref{8:15} must be rewritten, since the parameters $ \rho_{0} $ and $ R_{0} $ lose their meaning in the zero magnetic field of the diamagnetic bubble.

Referring to Fig.~\ref{fig:CoordinatesInCylinder}, the generalized azimuthal moment can be written as
    \begin{equation}
    p_{\psi} = mv_{\bot}r\sin(\theta - \psi) + \frac{e}{c}\,\chi(r,z)
    ,
    \end{equation}
similarly to \eqref{6:03a}, with the difference that $ v_{\bot} $ is the projection of the particle’s velocity onto a plane perpendicular to the axis of the diamagnetic bubble.


\begin{figure*}
  \centering
  \includegraphics[width=0.47\linewidth]{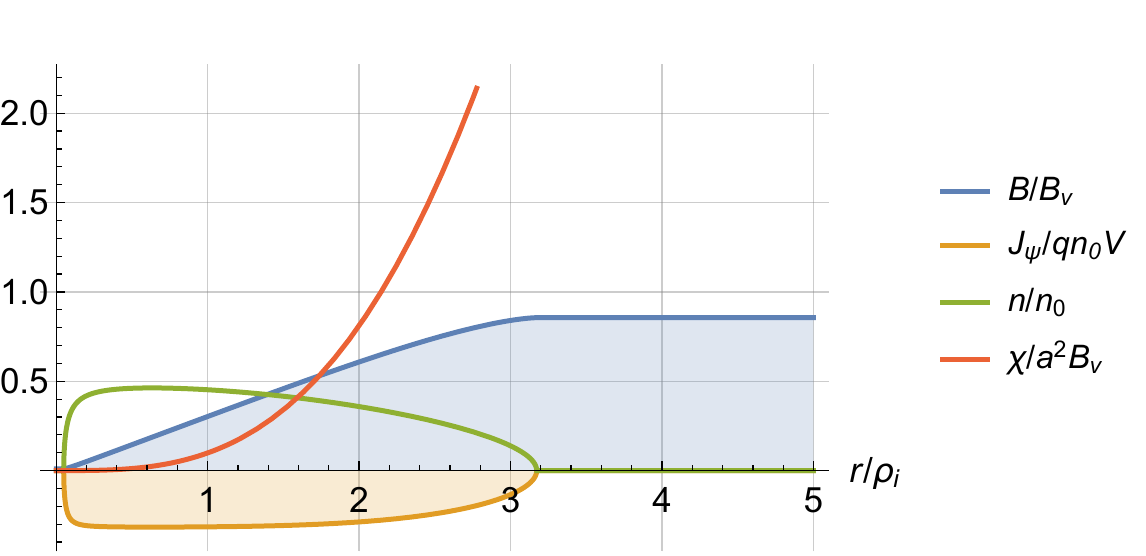}\hfil
  \includegraphics[width=0.47\linewidth]{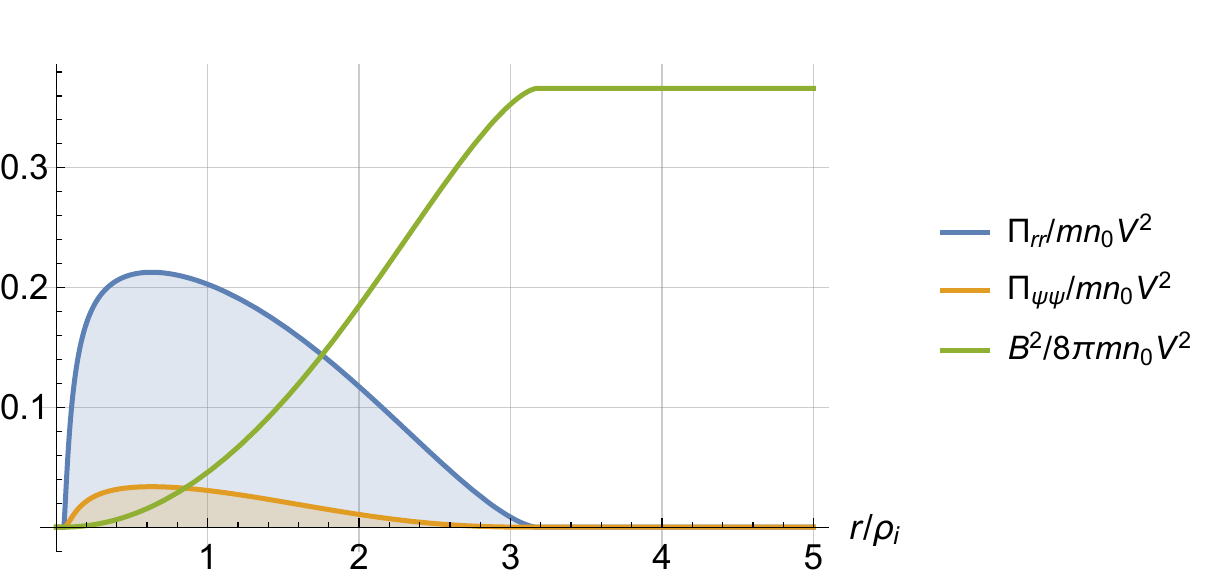}
  \caption{
    Radial profiles of the magnetic field, current density, density, magnetic potential (left) and components of the momentum flux tensor, magnetic field pressure (right) at $ B_{\textnormal{in}}/B_{\textnormal{v}} = 0.01 $ for case when the adiabatic invariant of ions is not conserved.
  }\label{fig:Abs(Bin,0_01)}
\end{figure*}

\begin{figure*}
  \centering
  \includegraphics[width=0.47\linewidth]{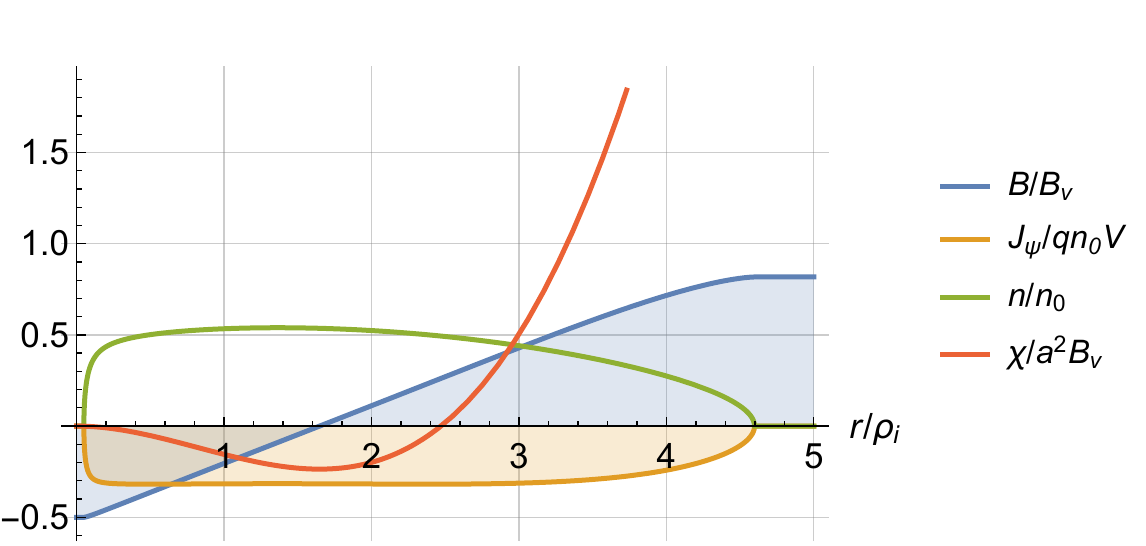}\hfil
  \includegraphics[width=0.47\linewidth]{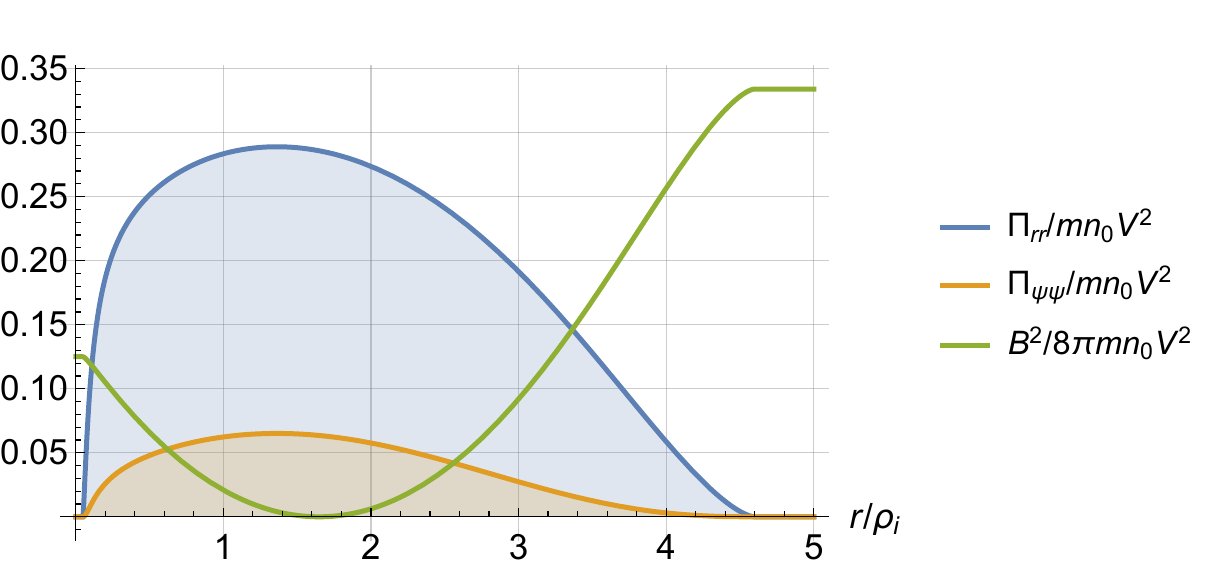}
  \caption{
    Radial profiles of the magnetic field, current density, density, magnetic potential (left) and components of the momentum flux tensor, magnetic field pressure (right) at $ B_{\textnormal{in}}/B_{\textnormal{v}} = - 0.5 $ for the case when the adiabatic invariant of ions is not conserved.
  }\label{fig:Abs(Bin,0_5)}
\end{figure*}
If we assume that due to a sharp change in the radius of the diamagnetic bubble near the magnetic plug, the adiabatic invariant is not preserved and therefore ions remain only in the absolute confinement region, their distribution function, by analogy with Eq.~\eqref{6:04}, can be written as
    \begin{multline}
    \label{6:04'}
    f(v,\theta)
    =
    \frac{n_{0}}{2\pi V}\,\delta (v-V)\,
    \times
    \\
    \Heviside\!\left(
        - \frac{v^{2}}{2\Omega_{\max}}
        -
        \left(r v\sin(\theta-\psi)+(e/mc)\,\chi\right)
    \right)
    ,
    \end{multline}
%
where $ \Omega_{\max} = eB_{\max}/mc $. Repeating further calculations similar to those described in Section \ref{s6}, it is possible to calculate the profiles of macroscopic quantities. The calculation results are shown in Fig.~\ref{fig:Abs(Bin,0_01)} and~\ref{fig:Abs(Bin,0_5)} for two values of the magnetic field $ B_ {\textnormal{in}} $ at $ r = 0 $. They indicate the disappearance of the internal cavity in the diamagnetic bubble in the sense that the magnetic field is everywhere inhomogeneous and noticeably different from zero. As can be seen from Fig.~\ref{fig:Abs(Bin,0_5)}, in this configuration it is possible to reverse the direction of the magnetic field in the axial region. In the immediate vicinity of the axis of the system, for $ r <V/2 \Omega_{\max} $, the plasma density, current density, and pressure are zero.


\section{Discussion and conclusions}\label{s99}

%

This article is devoted to the kinetic description of plasma equilibrium in the diamagnetic trap proposed by Alexey Beklemishev. The traditional description of the plasma behavior in such a trap using the drift theory is not applicable, since the ions move in a substantially non-circular orbit whose diameter is approximately equal to the diameter of the diamagnetic bubble. We found the ion distribution function in the collisionless approximation. Neglecting the diamagnetic current of electrons, which is permissible if their temperature is sufficiently low, we calculated the radial profile of the magnetic field, plasma density, current density, and pressure tensor. We found that the width of the boundary layer in the diamagnetic bubble varies from 6 to 8 Larmor radii evaluated for the vacuum magnetic field. In addition, we calculated the adiabatic invariant, replacing the magnetic moment, which is not preserved in the diamagnetic bubble. We formulated a criterion for absolute confinement and showed that a magnetic field penetrates into the diamagnetic bubble if the magnetic moment is not conserved and only ions in the absolute confinement region are trapped. This result means that ensuring the conditions for the conservation of the adiabatic invariant is critical to ensure the design mode of operation of the diamagnetic trap.


\begin{acknowledgements}

    The work was supported by the Ministry of Education and Science of the Russian Federation. The author is grateful to Vitaly Astrelin and Ivan Chernoshtanov for numerous fruitful discussions.

\end{acknowledgements}

\appendix
\section{Proof of conservation of the angle of incidence }\label{A1}


We study the change in the angle $ \alpha $ with an adiabatically slow change in the radius of the diamagnetic bubble.


As a preparatory example, we consider the problem of the motion of a ball between two slowly approaching walls. In a collision with the wall, the ball is elastically reflected. Let the left wall be motionless and the right one move away with the speed $ u $. If a ball hits this wall at a speed of $ V $, then it is reflected at the speed of $ - (V-2u) $, that is, the increment of speed over a period is $ \dif {V} = - 2u $. In this case, the distance $ L $ between the walls during the period will increase by $ \dif {L} = u \, (2L/V) $. It is easy to verify that the product $ VL $ is preserved in the first order at a low speed $ u $, that is, $ \left (V + \dif {V} \right) \left (L + \dif {L} \right) = VL + \mathcal {O} (u^{2}) $.

Let's try to repeat the same reasoning in the cylinder. Suppose that the cylinder radius $a$ slowly changes with the speed $ \dot {a} = u $. Let the particle move at a speed of $ V $ along the $ x $ axis at a distance of $ y $ from the cylinder axis. The angle $ \alpha $ is determined from the equations
    \begin{gather*}
    x=a\cos(\alpha),
    \qquad
    y=a\sin(\alpha ).
    \end{gather*}
At the start point of the particle  $ x^{2} + y^{2} = a^{2} $ but for the time $ \tau = 2a \cos (\alpha)/V $ before the next collision with the wall, the radius of the cylinder will change by the value $ \dif {a} = u \tau $. The coordinate $ y $ of the meeting point will still be equal to $ a \sin (\alpha) $, but the coordinate $ x $ will change by the value $ \dif {x} $, which can be determined from the equation
    \begin{equation*}
    (x+\dif{x})^{2} + y^{2} = (a+\dif{a})^{2}
    .
    \end{equation*}
Accordingly, the angle of incidence will change by a certain amount $ \dif {\alpha_{1}} $, which can be found from the equation
    \begin{equation*}
    \left(a+\dif{a}\right)\cos(\alpha +\dif{\alpha_{1}}\!)
    =
    x+\dif{x}
    .
    \end{equation*}
Assuming $ \dif {a} $, $ \dif {x} $, $ \dif {\alpha_{1}} $ to be small quantities of the same order, in the first order in $ u $ from these equations we find:
    \begin{gather*}
    \dif{a} = \frac{2au}{V}\,\cos(\alpha )
    ,\quad
    \dif{x} = \frac{2au}{V}
    ,\quad
    \dif{\alpha_{1}} = -\frac{2u}{V}\,\sin(\alpha )
    .
    \end{gather*}
In a collision with a moving cylinder wall, the tangential velocity
    \begin{equation*}
    V_{t}+\dif{V_{t}} = V \sin(\alpha +\dif{\alpha_{1}})
    \end{equation*}
is preserved, while the normal one receives an increment by $ -2u $:
    \begin{equation*}
    V_{n}+\dif{V_{n}} = V \cos(\alpha +\dif{\alpha_{1}}) - 2u
    .
    \end{equation*}
Therefore, the reflection angle $ \alpha + \dif {\alpha} $ will not be equal to the angle of incidence $ \alpha + \dif {\alpha_{1}} \! $. The total increment of the angle $ \dif {\alpha} $ and of total speed $ \dif {V} $ can be found from the equations
    \begin{gather*}
    \frac{ V_{t}+\dif{V_{t}} }{ V_{n}+\dif{V_{n}} }
    =
    \tan(\alpha +\dif{\alpha })
    ,\\
    (V_{t}+\dif{V_{t}})^{2} + (V_{n}+\dif{V_{n}})^{2}
    =
    (V+\dif{V})^{2}
    .
    \end{gather*}
Solving these equations up to linear corrections in $ u $, we obtain:
    \begin{gather*}
    V_{t} = V \sin(\alpha )
    ,\qquad
    V_{n} = V \cos(\alpha )
    ,\\
    \dif{V_{t}} = -u \sin(2\alpha )
    ,\qquad
    \dif{V_{n}} = -2u \cos^{2}(\alpha )
    ,\\
    \dif{V} = -2u \cos(\alpha )
    ,\qquad
    \dif{\alpha } = 0.
    \end{gather*}
It follows from them that the angle $ \alpha $ and the product $ Va $ are adiabatic invariants.



%

\end{document}